\documentclass[useAMS]{mn2e}

\def\var{\hbox{V\,393\,Sco~}}

\def\op{\hbox{$\Phi_{o}$ =~}}
\def\lp{\hbox{$\Phi_{l}$ =~}}
\def\kms{\hbox{km s$^{-1}$}}

\usepackage{graphicx}
 \usepackage{times}
 \usepackage{lscape}

\title{A cyclic  bipolar wind  in the interacting binary V\,393 Scorpii}
\author[ Mennickent et al.]
  {R.E. Mennickent$^{1}$\thanks{E-mail: rmennick@astro-udec.cl. Based on ESO proposal 081.D-0457(B)} ,   
  Z. Ko{\l}aczkowski$^{2}$,  
G. Djurasevic$^{3,4}$,
    E. Niemczura$^{2}$,
    M. Diaz$^{5}$, \newauthor
    M. Cur\'e$^{6}$, 
    I. Araya$^{6}$,
    G. J. Peters$^{7}$
 \\   
  $^1$Universidad de Concepci\'on, Departamento de Astronom\'{\i}a,
      Casilla 160-C, Concepci\'on, Chile\\
 $^2$ Instytut Astronomiczny Uniwersytetu Wroc{\l }awskiego
Ul. Kopernika 11, 51-622 Wroc{\l }aw, Poland\\
  $^{3}$Astronomical Observatory, Volgina 7, 11060 Belgrade 38, Serbia\\
   $^{4}$ Isaac Newton Institute of Chile, Yugoslavia Branch\\
  $^{5}$ Departamento de Astronomia, IAG,
U. de Sao Paulo, Rua do Matao, 1226, Butanta
05508-900, Sao Paulo, SP, Brasil\\
$^{6}$ Departamento de F\'{\i}sica y Astronom\'{\i}a, Facultad de Ciencias, Universidad de Valpara\'{\i}so, Chile\\
$^{7}$ Space Sciences Center/Department of Physics \& Astronomy, University of Southern California, Los Angeles, CA 90089-1341, USA
  }
\date{}

\pagerange{\pageref{firstpage}--\pageref{lastpage}} \pubyear{2008}

\def\LaTeX{L\kern-.36em\raise.3ex\hbox{a}\kern-.15em
    T\kern-.1667em\lower.7ex\hbox{E}\kern-.125emX}

\begin{document}

\label{firstpage}

\maketitle

\begin{abstract} 

V\,393 Scorpii is  a Double Periodic Variable characterized by a relatively stable  non-orbital   photometric cycle of 253 days. Mennickent et al. argue for the presence of a massive optically thick disc around the  more massive B-type component and describe the evolutionary stage of the system. In this paper we analyze the behavior of the main  spectroscopic optical lines during the long non-orbital photometric cycle. We study the radial velocity of the donor determining their orbital elements and find a small but significant orbital eccentricity ($e$ = 0.04). The donor spectral features are modeled and removed from the spectrum at every observing epoch using the light-curve model given  by Mennickent et al.  We find that the line emission is larger during eclipses and mostly comes from a  bipolar wind.  We find that the long cycle is explained in terms of a modulation of the wind strength; the wind has a larger line and continuum emissivity on the high state. We report the discovery of highly variable chromospheric emission  in the donor, as revealed by Doppler maps of the emission lines Mg\,II\,4481 and C\,I\,6588.
We discuss notable and some novel spectroscopic features like discrete absorption components, especially visible at blue-depressed O\,I\,7773 absorption wings during the second half-cycle, Balmer double emission with $V/R$-curves showing ``Z-type`` and ``S-type'' excursions around secondary and main eclipse, respectively, and H$\beta$ emission wings extending up to $\pm$ 2000 km s$^{-1}$. We discuss possible causes for  these  phenomena and for  their modulations with the long cycle.

\end{abstract}


\section{Introduction}
 
V\,393 Scorpii is one of the Galactic  Double Periodic Variables (DPVs),  a group of interacting binaries showing a long photometric cycle lasting roughly 33 times  the orbital period (Mennickent et al. 2003, 2005, 2008, Mennickent \& Ko{\l}aczkowski 2010,  Radek et al. 2010). DPVs  have been interpreted as semi-detached interacting binaries  ongoing cyclic episodes of mass loss into the interstellar medium  (Mennickent et al. 2008, Mennickent \& Ko{\l}aczkowski 2010). The long photometric cycle of V\,393\,Sco (253 days) was discovered by Pilecki \&  Szczygiel (2007)  by inspecting the ASAS catalogue of eclipsing binaries with additional variability. 
Ultraviolet spectra of V\,393 Sco were studied by Peters (2001) who found evidence for a hot temperature region that should be the origin of superionized lines like N\,V, C\,IV and Si\,IV. These lines are likely produced by resonance scattering in a plasma of  temperature T $\sim$ 10$^{5}$ K and electron density $N_{e} \sim$  10$^{9}$ cm$^{-3}$ (Peters \& Polidan 1984).  Broad band photometry was modeled by Mennickent et al. (2010a, hereafter M10), who found  a distance $d$ = 523 pc and reddening $E(B-V)$ = 0.13. They  interpreted  observed asymmetries in UV lines of highly ionized metals as evidence for a high-latitude wind. They also studied orbitally-resolved high resolution infrared spectra and found a remarkable depressed blue wing in the He\,I\,1083 nm line near secondary eclipse, that interpreted as evidence of mass loss through the $L3$ point. The relatively stable ultraviolet  spectra during the long cycle yielded these authors to suggest that the mass outflow related to the long cycle should be through the equatorial plane.

Posteriorly, Mennickent et al. (2012, hereafter M12) disentangled the orbital and long-cycle light curves and performed a multi parametric fit to the orbital light curve considering two stellar components plus a circumprimary disc.  The best fit is obtained with a Roche-lobe filling 
A7 giant  (hereafter the donor or secondary star) and a B4 dwarf (the gainer or primary) that  is surrounded by an optically thick  disc with a radius almost twice the radius of its central star. The disc has 16.600 $K$ in the inner parts and 8.600 $K$ in the outer regions. The orbit is circular with separation  35 $R_{\sun}$ and the inclination is 80 degree. M12 found two bright spots in the disc, that are needed to fully reproduce the shape of the orbital light curve. These spots are 20\% hotter that the disc and face the observer at orbital phases 0.90 and 0.45.

M12 performed a $\chi^{2}$ minimization  procedure to find the system and orbital parameters that best match a grid of theoretical evolutionary tracks for binary stars including non-conservative processes. After performing this procedure, M12 find an age of 70 Myr, but more importantly, they found that the system is just after a mass transfer burst that transferred 4 $M_{\sun}$ from the donor to the gainer in just 400.000 years. The system and orbital parameters are reproduced relatively well, but a mismatch between the gainer dynamical mass and its luminosity is found. In order to explain this discrepancy, M12 argue that not all transferred mass has been accumulated by the (possibly critically rotating) gainer, but a significant fraction  ($\sim$ 2 M$_{\odot}$) still remains in the disc. Regarding the long cycle, M12 find that  {\it is not} produced by changes in the relatively stable disc, and invoke a variable high latitude wind as probable cause for the phenomenon. Emission in this wind should explain, according to them, the non-eclipsing nature of the long variability and their red colors at maximum.


With the aim of studying the nature and origin of the long photometric cycle, we carried out a spectroscopic campaign of V\,393 Sco using several high resolution spectrographs at different observatories in Chile. Additional spectra taken 20 years ago at Kitt Peak National Observatory  were also considered. The analysis of  the main spectral features of V\,393\,Sco  is presented in this paper. We postpose for a future paper 
the analysis of the weaker spectroscopic features.

In this paper long-cycle maximum and minimum are indistinctly named high and low state, respectively. We use the ephemerides for the orbital and long cycles given by M12, viz.\,, the main orbital photometric minimum happens at $HJD_{o} = 2\,452\,507.7800 + 7.7125772 E$ (Kreiner 2004) and the maximum of the long cycle occurs at $HJD_{l} = 2452522 + 253E$.   The observations, instruments and reduction techniques are presented in Section 2, the results are given in Section 3, a general discussion is presented in Section 4 and the conclusions  are given in Section 5. A brief preliminary report of this research was  published in a conference proceeding by Mennickent et al. (2010b).

\section{Spectroscopic observations and data reduction}

We conducted spectroscopic observations of V\,393\,Sco during 2 years  obtaining optical  spectra with resolution  $R \sim$ 40.000 with several echelle spectrographs in Chile. The instruments used were CORALIE, HARPS and FEROS (La Silla ESO Observatory, 106, 6 and 14 spectra, respectively), UVES (Paranal ESO Observatory, 359 spectra) and DuPont-echelle and MIKE (Las Campanas Observatory, 42 and 17 spectra, respectively)\footnote{Technical descriptions for these spectrographs and their cameras can be found in www.lco.cl/ , www.eso.org/sci/facilities/paranal/instruments/ and www.eso.org/sci/facilities/lasilla/instruments/.}.
The studied spectral regions were 3865-6900 \AA~ (CORALIE), 3780-6910 \AA~ (HARPS), 3560-9215 \AA~ (FEROS), 3600-9850 \AA~  (DuPont-echelle) 3350-5000 \& 4900-9500 \AA~ (MIKE), 3760-4985,  5682-7520 \& 7660-9464 \AA~ (UVES). 
We also included  9 medium resolution ($R \approx$ 15.000) spectra in the H$\alpha$ region and 2 in the He\,4471 region  ($R \approx$ 22.000) obtained between years 1987-1990 with the COUDE spectrograph in the Kitt Peak National Observatory. Since the long-cycle phases are quite uncertain these spectra are only phased with the orbital cycle. In general, their variability is not different from spectra acquired 20 years later. A total of 555 spectra were acquired during our campaign, most of them with signal to noise ratio larger than 100. The histograms for the orbital and long-cycle phases of the high resolution spectra obtained in the  H$\alpha$ region show that our observations cover relatively well the short and long periodicity  (Fig.\,1).


All spectra discussed in this paper are corrected by earth translational motion and normalized to the continuum. 
The RVs are heliocentric ones. Reductions were done with IRAF\footnote{IRAF is distributed by the National Optical Astronomy Observatories,
 which are operated by the Association of Universities for Research
 in Astronomy, Inc., under cooperative agreement with the National
 Science Foundation.} following usual procedures for echelle spectrography, including flat and bias correction, wavelength calibration and order merging. Occasional cleaning of spectra was needed due to the presence of bright features produced by incident cosmic rays or the presence of sharp telluric lines. In order to remove these features, we replaced deviant points from a high order fit to the flux distribution in a short spectral segment with the flux corresponding to the fitting function.
Most cosmic ray features  and telluric lines were efficiently removed with this technique.

Some of the DuPont spectra showed artifacts around short spectral regions produced by interference of light reflected in  the detector inner layers. These ``fringes'' were identified and the fluxes in the affected regions were interpolated and not considered for the analysis. 

The spectra obtained with the CORALIE spectrograph are not sky-subtracted.  This condition imposed by the instrumental setup  has not effect for radial velocity  and  line strength measurements, since V\,393\,Sco is very bright even at full moon and we do not flux-calibrate our spectra.
Details for our observational runs  are given in Tables 1, 2 and 3. 


\begin{table*}
\centering
 \caption{Summary of LCO and ESO non-UVES spectroscopic observations. N is number of spectra. The HJD at mid-exposure for the first spectrum of the series is given. $\Phi_{o}$ and $\Phi_{l}$ refer to the orbital and long-cycle phase, respectively.}
 \begin{tabular}{@{}cccccrcc@{}}
 \hline
UT-date &Observatory/Telescope & Instrument &N & exptime (s)  &HJD-MID (start) &$\Phi_{o}$& $\Phi_{l}$\\
\hline
2007-07-19&LCO/Magellan Clay & MIKE & 1  &20               &2454301.47395  &0.5674  & 0.0335 \\
2007-07-20&LCO/Magellan Clay  & MIKE & 6 &230            &2454301.50920  &0.5720  & 0.0336 \\
2007-07-21&LCO/Magellan Clay  & MIKE   & 4  &40       &2454302.53377  &0.7048  & 0.0377\\
2007-08-01&LCO/Magellan Clay  & MIKE    & 1  & 920        &2454313.62004  &0.1422  &  0.0816\\
2007-08-02&LCO/Magellan Clay  & MIKE    & 1  & 390        &2454314.55761  &0.2638  & 0.0852\\
2007-08-05&LCO/Magellan Clay  & MIKE    & 1  & 250        &2454317.70700  &0.6721  & 0.0977\\
2007-08-16&LCO/Magellan Clay  & MIKE   & 1  & 270        &2454328.58438  &0.0825  &  0.1406\\
2007-11-08&LCO/Magellan Clay  & MIKE   & 1  &120        &2454412.50750  &0.9638  & 0.4724\\
2007-11-09&LCO/Magellan Clay  & MIKE   & 1  &160         &2454413.49772  &0.0922  & 0.4763\\
2008-02-24&LCO/Du Pont & Echelle & 8 &500    &2454520.83187        &0.0090  &0.9005   \\
2008-02-25&LCO/Du Pont & Echelle & 6 &500    &2454521.83514        & 0.1391 &0.9045\\
2008-06-15&LCO/Du Pont & Echelle & 11 &500 &2454632.51366        &0.4895  &0.3420\\
2009-05-14&LCO/Du Pont & Echelle & 7  &420  &2454965.65937        &0.6846  &0.6587\\
2009-06-16&LCO/Du Pont & Echelle & 6  &540--600 &2454998.63019        &0.9595  &0.7891\\
2009-08-25&LCO/Du Pont & Echelle  & 4 &540&2455068.57889 &0.0289       &0.0655\\
2009-02-26&ESO/ 3.6m &HARPS&1 &300 &2454888.87719   &0.7291    &0.3552\\
2009-02-28&ESO/ 3.6m &HARPS&1 &180 &2454890.82302    &0.9814   &0.3629\\
2010-09-09&ESO/ 3.6m &HARPS&1 &180  &2455448.51123     &0.2900     &0.5672\\
2010-09-27&ESO/ 3.6m &HARPS&1 &180   &2455467.47358     &0.7480     &0.6422\\
2010-09-28&ESO/ 3.6m &HARPS&1 &210   &2455468.48994     &0.8800     &0.6462\\
2010-09-30&ESO/ 3.6m &HARPS&1 &250  &2455470.47588     &0.1380     &0.6541\\
2008-03-16&ESO/2.2m & Feros  & 3 &250--300 &2454541.89607&0.7401 &0.9838\\
2008-03-17&ESO/2.2m  & Feros  &  3 &300--400 &2454542.90141&0.8705 &0.9878\\
2008-03-18&ESO/2.2m  & Feros &  3 &300--400 &2454543.89221&0.9989 &0.9917\\
2008-03-19&ESO/2.2m  & Feros &  2 &500           &2454544.89839&0.1294 &0.9956\\
2008-03-20&ESO/2.2m  & Feros &  3 &400--500 &2454545.89879&0.2591 &0.9996\\
2008-04-06&ESO/EULER &CORALIE  &6 &900&2454562.82551    & 0.4538 &0.0665\\
2008-04-07&ESO/EULER &CORALIE  &8 &1000&2454563.77105  &0.5764  &0.0702\\
2008-04-08&ESO/EULER &CORALIE  &4 &900&2454563.77105    & 0.5764 &0.0702\\
2008-04-09&ESO/EULER &CORALIE  &4 &1000&2454565.83190  &0.8436  &0.0784\\
2008-04-10&ESO/EULER &CORALIE  &1 &900&2454566.92395    & 0.9852 &0.0827\\
2008-05-23&ESO/EULER &CORALIE  &2&1000 &2454609.84474  &0.5502  &0.2524 \\
2008-05-24&ESO/EULER &CORALIE  &4&1000 &2454610.78428  &0.6721  &0.2561\\
2008-05-25&ESO/EULER &CORALIE  &5&1000&2454611.80897   &0.8049  &0.2601\\
2008-05-26&ESO/EULER &CORALIE  &7&1000&2454612.77847  & 0.9306 &0.2639\\
2008-08-19&ESO/EULER &CORALIE  &1 &600&2454698.53568 &   0.0498&   0.6029\\
2008-08-20&ESO/EULER &CORALIE  &1 &600&2454699.48964 &   0.1734&   0.6067\\
2008-08-21&ESO/EULER &CORALIE  &1 &600&2454700.47897 &   0.3017&   0.6106\\
2008-08-22&ESO/EULER &CORALIE  &2 &600&2454701.63242 &   0.4513&   0.6151\\
2008-08-23&ESO/EULER &CORALIE  &2 &600&2454702.64402 &   0.5824&   0.6191\\
2008-08-24&ESO/EULER &CORALIE  &2 &600&2454703.59256 &   0.7054&   0.6229 \\
2008-08-25&ESO/EULER &CORALIE  &1 &600&2454704.48019 &   0.8205&   0.6264\\
2008-08-26&ESO/EULER &CORALIE  &1 &600&2454705.50720 &   0.9537&   0.6305\\
2008-08-28&ESO/EULER &CORALIE  &1 &600&2454707.51773 &   0.2144& 0.6384\\
2008-10-01&ESO/EULER &CORALIE  &6 &600--800  &2454741.53378&    0.6248&   0.7729\\
2008-10-02&ESO/EULER &CORALIE  &3 &900--1000&2454742.52995&    0.7540&0.7768\\
2008-10-03&ESO/EULER &CORALIE  &4 &900  &2454743.52755&    0.8833&  0.7807\\
2009-04-15&ESO/EULER &CORALIE  &1 &1800&2454936.90276&    0.9560&  0.5451\\
2009-04-16&ESO/EULER &CORALIE  &3 &1300&2454937.87533&    0.0821&  0.5489\\
2009-04-17&ESO/EULER &CORALIE  &4 &1000&2454938.87689&    0.2120&  0.5529 \\
2009-05-17&ESO/EULER &CORALIE  &11 &900&2454968.61926&    0.0683&  0.6704\\
2009-05-18&ESO/EULER &CORALIE  &9 &900  &2454969.72474&    0.2117&  0.6748\\
2009-05-19&ESO/EULER &CORALIE  &12 &800&2454970.59597&    0.3246&  0.6782\\
\hline
\end{tabular}
\end{table*}

\begin{table}
\centering
 \caption{Summary of KPNO spectroscopic observations obtained with the Coude spectrograph. One spectrum was obtained per night. $\Phi_{o}$ is the orbital phase.}
 \begin{tabular}{@{}ccccc@{}}
 \hline
UT-date  &$\Delta \lambda$ (\AA)  & exptime (s)  &HJD &$\Phi_{o}$\\
\hline
1987-05-03& 6520-6695  &1800 &2446918.93455 &0.3595  \\
1987-08-24&  6520-6695 &2700 &2447031.69913&0.9804  \\
1989-04-20&  6520-6695  &2400 &2447636.97037&0.4589  \\
1989-04-21&  6520-6695 &2700 &2447637.92552&0.5827   \\
1989-04-23&  6520-6695 &2700 &2447639.97292& 0.8482  \\
1989-04-24&  6520-6695  &3300 &2447640.94252& 0.9739  \\
1990-08-19&  6520-6695 &2000 &2448122.68538& 0.4359  \\
1990-08-20&  4420-4535 &2700 &2448123.65323& 0.5614  \\
1990-08-21& 4420-4535 &3600  &2448124.70935 &  0.6983\\
1990-08-22& 6520-6695  &1800 &2448125.65894&  0.8214 \\
1989-94-21& 6520-6695  &2700 &2447637.92552&  0.5827 \\
\hline
\end{tabular}
\end{table}

\section{Results}

\subsection{General spectrum appearance and interstellar reddening}

The optical spectral region  of V\,393 Sco shows Balmer, He\,I, Si\,II lines  in absorption and sometimes strong emission in  H$\alpha$, suggesting a B-type primary surrounded by a variable gaseous envelope, along with metallic absorption lines, interpreted as signatures of the A-type secondary star.  A rapid inspection showed several absorption helium lines varying  in width and depth, a fact that is consistent with a non-photospheric origin. The above description suggests the presence of circumstellar matter and it is consistent with the earlier classification of intermediate-mass interacting binary for this system.

The interstellar contribution to the color excess can be determined through the analysis of diffuse interstellar bands (DIBs; Munari 2000, Weselak et al. 2008). We measured equivalent widths of 
DIBs located at    5780 \AA, 5797 \AA~ and 8620 \AA , estimating $E(B-V)$ =  0.15 $\pm$ 0.05, in agreement with the value $E(B-V)$= 0.13 $\pm$ 0.02 obtained by M10 from the fit to the spectral energy distribution. 
The above probably excludes significant amounts of dust in the line of sight to V\,393 Scorpii and suggests that the color excess is mainly due to interstellar reddening rather than circumstellar matter.



\subsection{The donor's spectrum and its stellar parameters}

In order to find the donor spectral type we compared the spectrum observed at minimum light with a grid of synthetic spectra and looked for the best match in a region 
dominated by donor spectral features. For the reference spectrum 
we used the FEROS spectrum  at $\Phi_{o}$ =  0.000 obtained on March 18, 2008 (hereafter $\Phi_{o}$ and $\Phi_{l}$ stand for orbital and long cycle phase respectively).
The region  used for the fit, deployed of hydrogen and helium lines but with several metallic lines, was the region between 4150 and 4290 \AA.

To determine the grid of synthetic fluxes we used atmospheric models computed with the line-blanketed LTE ATLAS9 code (Kurucz  1993), which treats line opacity with opacity distribution functions (ODFs). The Kurucz's models are constructed with the assumptions of plane-parallel geometry and hydrostatic and radiative equilibrium of the gas. The synthetic spectra were computed with the SYNTHE code (Kurucz 1993). Both codes, ATLAS9 and SYNTHE were ported under GNU Linux by Sbordone et al. \cite{sbordone} and are available online\footnote{wwwuser.oat.ts.astro.it/atmos/}. The atomic data were taken from Castelli \& Hubrig \cite{castelli}\footnote{http://wwwuser.oat.ts.astro.it/castelli/grids.html}. The theoretical models were obtained for effective temperatures from 6000 to 10000 $K$ with steps of 100 $K$ and for surface gravities from 2.0 to 4.5 dex with the step of 0.1 dex. Solar and 0.5 dex higher metallicities were taken into account. The grid of synthetic spectra was calculated for five different rotation velocities, $v_{2r} sin i$ = 0, 25, 50, 75 and 100 \kms.


We subtracted the template from every grid spectrum, allowing small   wavelength adjustments and analyzed the residuals of the resulting difference spectra. The model with the lower root mean square had $T_{2}$ = 7900 $K$, $\log g_{2}$ = 3.0, $v_{2r} sin\,i =$ 75 km s$^{-1}$ and
solar metallicity.  Several solar metallicity models with temperatures in the range 7500-8100 $K$ showed similarly small residuals. Some higher metallicity models also fit the spectrum but with higher temperatures, the best one with 8400 $K$,  $\log g$ = 3.2, and $v_{2r} sin\,i$ = 75 km s$^{-1}$.  We preferred the lower temperature model for consistency with our earlier calculation and previously published results (M10 and references therein).

\begin{table}
\centering
 \caption{Summary of observations obtained with the ESO UVES between April 02 and October 01, 2008. N is the number of spectra.
The exposure time was 35 s. The complete table is given in electronic form.}
 \begin{tabular}{@{}ccc@{}}
 \hline
 N&$\Delta \lambda$ &HJD range \\ 
 & (\AA)  & (2450000 +)  \\ \hline
119&3760-4985&4558.90352-4740.54141\\
120&5682-7520& 4558.90350-4740.54142\\
120&7660-9464&4558.90350-4740.54142\\
\hline
\end{tabular}
\end{table}

\begin{figure}
\scalebox{1}[1]{\includegraphics[angle=0,width=8.5cm]{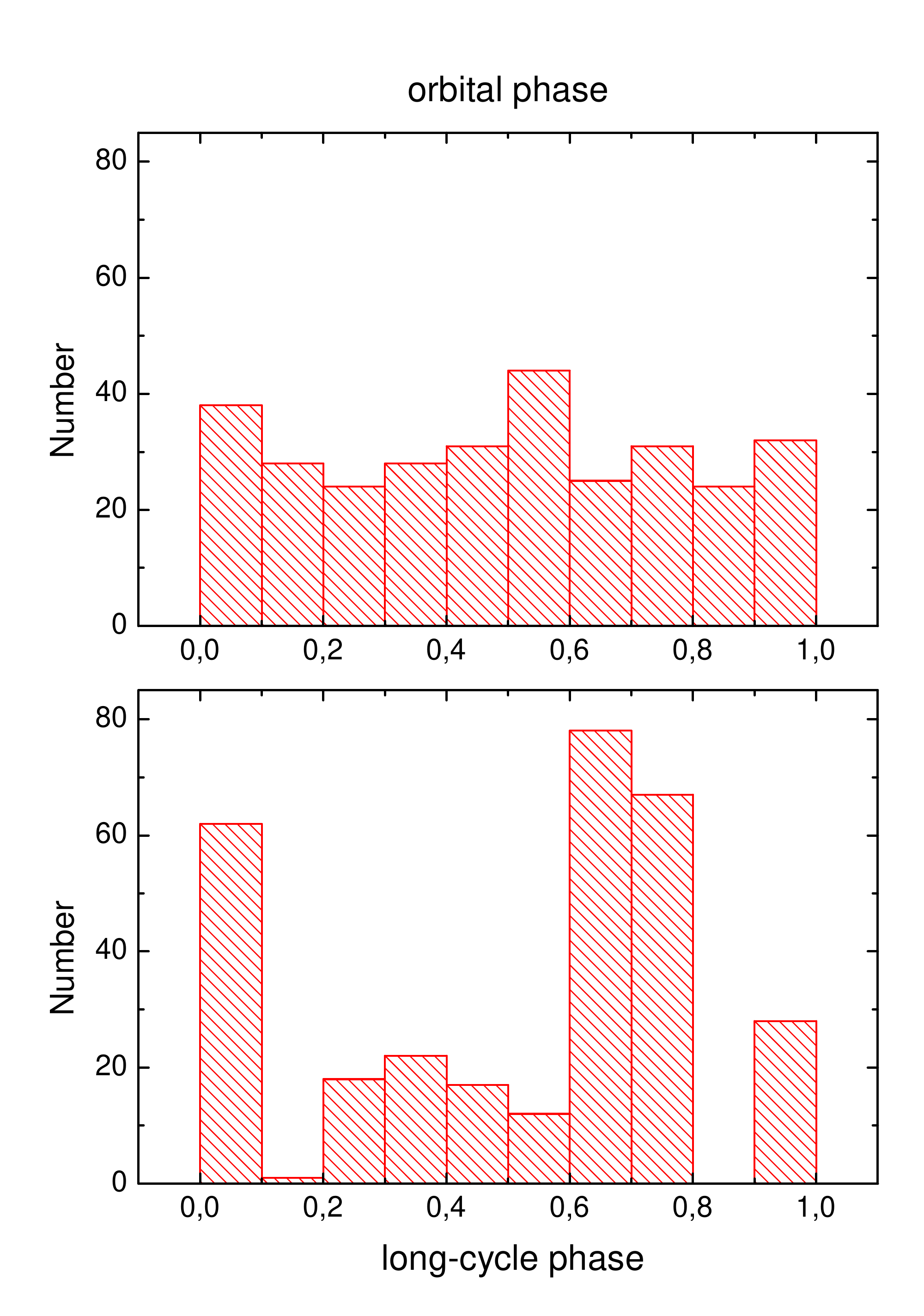}}
\caption{Number of H$\alpha$ spectra per bin of long  and orbital phase.}
  \label{x}
\end{figure}

\begin{figure*}
\scalebox{1}[1]{\includegraphics[angle=0,width=18cm]{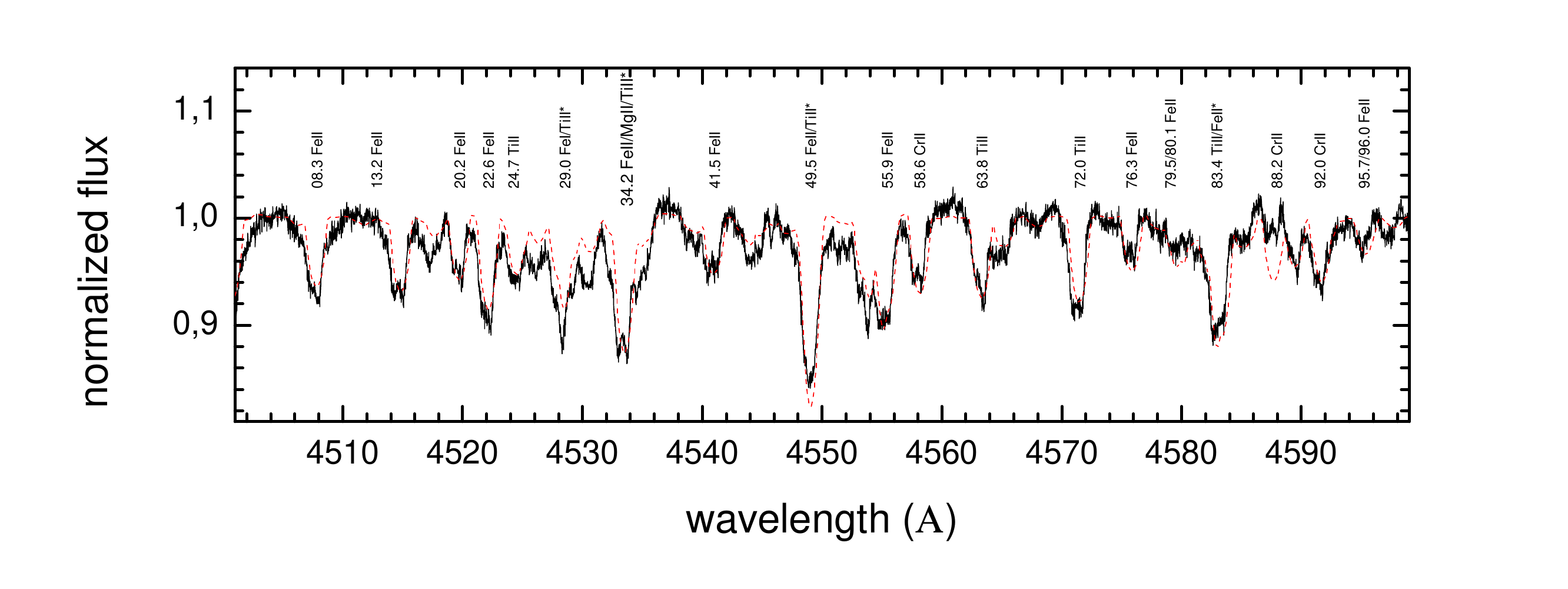}}
\caption{The donor synthetic spectrum (DSS, dashed line) over-plotted with one of our UVES spectrum taken near main eclipse (\op 0.972, \lp 0.753). To get the DSS, the model synthetic spectrum has been scaled according to the light curve model (M12) to fit the percentage of light contributed by the donor at this epoch. It has also been shifted in wavelength  to fit the donor radial velocity at this orbital phase. In the figure, the DSS is vertically shifted to match its continuum with the unity. Metallic lines were identified and labeled  ($\lambda$ $-$ 4500 \AA); uncertain identifications are marked with asterisks.}
  \label{x}
\end{figure*}

From the above we estimate  $T_{2}$ = 7900 $^{+200}_{-400}$ $K$. Our low temperature limit is consistent with the absence of the G-band in our spectra. The low sensitivity of spectral features  to the donor temperature  in the chosen spectral region precludes a better determination of $T_{2}$. Other potential temperature spectral diagnostics like  calcium H \& K lines could not be used due to contamination with features of other system components.

A refinement of $v_{2r} sin\.i$ was done by interpolating the grid of models to the $FWHM$ of 
lines observed on main eclipse, obtaining  $v_{2r}sin\,i$ = 60 $\pm$ 2 km s$^{-1}$. A new synthetic model for the donor was constructed using this rotational velocity. This model is used in the rest of this paper. The donor synthetic spectrum  compared with a typical UVES spectrum reveals in general a good match except for minor deviations in short wavelength segments due to difficult normalization of the waving continuum (Fig.\,2).


Looking at luminosity-sensitive features like the Fe II \& Ti II double blend at  4172-8 \AA, and similar blends at 4395-4400 \AA, 4417 \AA~ and 4444 \AA, we do not observe evidence  for  systematic changes in luminosity for the donor star. The same is true for its temperature, as indicated by line strength ratios.




\begin{figure}
\scalebox{1}[1]{\includegraphics[angle=0,width=8.5cm]{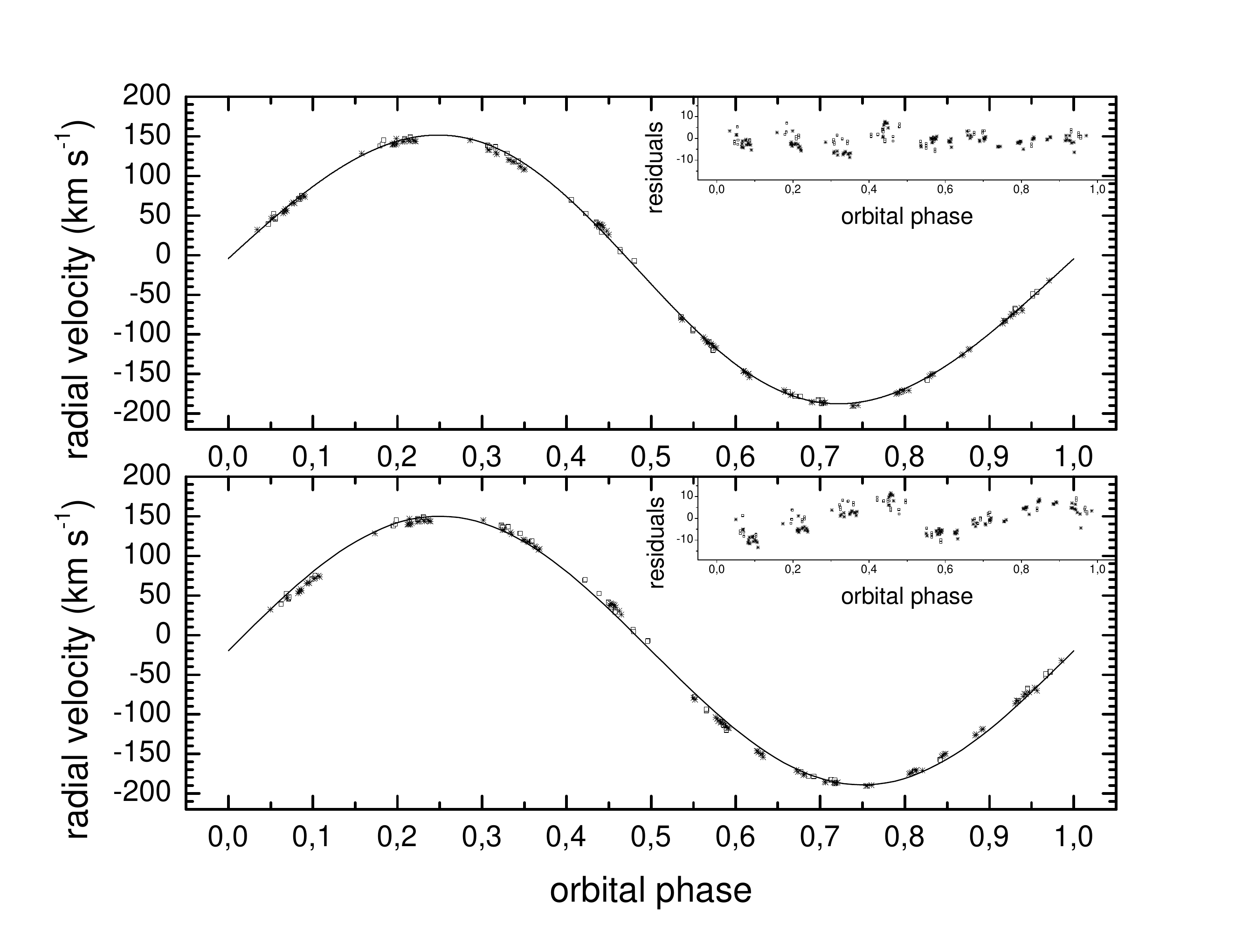}}
\caption{Radial velocities of the donor star  (asterisks for CORALIE data  and squares for UVES data) and the best fit for elliptical (up) and circular (down) orbits. Residuals are given in the inset graphs. 
 }
  \label{x}
\end{figure}

\begin{table}
\centering
 \caption{Donor radial velocities obtained by cross-correlation. Labels indicate UVES (U) and CORALIE (C) data. Typical RV error is 1 \kms. The complete table is given in electronic form.}
 \begin{tabular}{@{}crc@{}}
 \hline
HJD& RV (\kms) &label \\ 
 \hline
2454558,90258 &-55,62 &U \\
2454558,90352 &-54,56 &U \\
2454558,90446 &-55,59 &U \\
2454559,85331 &65,31 &U \\
2454559,85425 &65,31 &U\\
2454560,85597 &157,97 &U\\
2454560,85691 &158,01 &U\\
2454560,85784 &157,95 &U\\
2454561,87559 &149,95 &U\\
2454561,87653 &149,97 &U\\
\hline
\end{tabular}
\end{table}

\subsection{Radial velocities for the donor}

Radial velocities for the donor were measured by cross-correlating the observed spectra with the FEROS spectrum taken at mid eclipse. 
After obtaining relative velocities with respect to this reference spectrum, we applied a zero-point shift considering the template RV, measured by fitting the position of a set of known lines with simple gaussians and comparing these wavelengths with the corresponding laboratory wavelengths.
The cross-correlation was performed in  two regions deployed of H\,I and He\,I lines, viz.\,4500--4800 \AA~ and 5050--5680 \AA.  The radial velocities are given in Table\,4. 

The RVs can be fitted with a sinusoid with radial-velocity  half-amplitude $K_{2}$ = 169.7 $\pm$ 0.6 km s$^{-1}$ and  zero point  $\gamma$ = -19.6 $\pm$ 0.4 km s$^{-1}$.  These figures are significantly different from $K_{2}$ = 181 $\pm$ 3 and  $\gamma$ =  -2 $\pm$ 2 km s$^{-1}$ obtained from the analysis of the single line Mg\,II\,10917 by M10, but this line
is probably partly blended  with Pa$\gamma$ and He\,I\,10916 and affected by  residual emission. 
A careful inspection of Fig.\,3 shows non-random residuals for the circular fit. 


 In order to resolve the question about the possible ellipticity of the orbit, we used the genetic algorithm PIKAIA developed by Charbonneau (1995) to find the orbital elements for
V\,393\,Sco. The method consists in finding the set of orbital parameters that produces a series of theoretical velocities that minimize the  function $\chi^{2}$ defined as:\\

\begin{small}
$\chi^{2} (P_{o}, \tau, \omega, e, K_{2}, \gamma) = $\\

$\frac{1}{N-6} \displaystyle\sum\limits_{j=1}^N 
(\frac{V_{j}-V(t_{j}, P_{o}, \tau, \omega, e, K_{2}, \gamma)}{\sigma_{j}})
$\hfill(1)\\
\end{small}

\noindent
where $N$ is the number of observations, $P_{o}$ is the orbital period, $\omega$ the periastron longitude, $\tau$ the time of passage per the periastron, $e$ the orbital eccentricity, $K_{2}$ the half-amplitude of the radial velocity for the donor and $\gamma$ the velocity of the system center of mass. $V_{j}$ and $V$ are the observed and theoretical radial velocities at $t_{j}$. The theoretical velocity is given by:\\

$V(t) = \gamma + K_{2} (cos (\omega+v(t)) + e~ cos (\omega))$\hfill(2)\\

\noindent
where $v$ is the angular velocity around the system center of mass obtained resolving the following two equations involving the eccentric anomaly $E$:\\

$tan (\frac{v}{2}) = \sqrt{\frac{1+e}{1-e}}~ tan (\frac{E}{2}) $\hfill(3)\\

$E - E~sin(E) = \frac{2 \pi}{P_{o}} (t - \tau) $\hfill(4)\\

A range of physically reasonable parameters need to be considered so that the method works. For the period we used the range 0-10 days, the eccentricity was set between 0 and 1, $\omega$ between 0 and $2\pi$, $\tau$ between the minimum julian day and this value plus $P_{o}$, $K_{2}$ between 0 and ($V_{max} - V_{min}$) and $\gamma$ between $V_{min}$ and $V_{max}$. 

The most reliable way to get error estimates for this genetic algorithm is by  Monte Carlo simulations, specifically by perturbing the best fit solution and computing the
$\chi^{2}$ of these perturbed solutions. To find the standard deviation region ($\sigma$)
 encompassed by the joint variation of two parameters with all other
parameters at their optimized values, we draw the contour corresponding to that value
of $\Delta \chi^2$ for 2 degrees of freedom that includes 68.3\% of the
probability. In our case this corresponds to $\Delta \chi^2 = 2.30$ 
(Bevington \& Robinson 1992, Chapter 11, pp. 212).

The best orbital elements for the circular and eccentric solutions along with their estimated errors are presented in Table 5. It is clear from the $\chi^{2}$ value in this table and the residual plots in Fig.\,3 that the 
elliptical solution provides the best fit, since it gives residuals without systematic trends and also the smaller  $\chi^{2}$ value. We note that our eccentric solution gives a small $e$ value (0.042) but it is highly significant, according to the statistical test ``$p_{1}$'' of Lucy (2005).

It has been pointed out that gas stream and circumstellar matter can distort spectroscopic features in semi-detached interacting binaries, producing skewed radial velocities and artificial small eccentricities (e.g. Lucy 2005). For a non-interacting binary with the stellar and orbital parameters of V\,393\,Sco, dynamical tides should have circularized the orbit and synchronized the rotational  periods (Zahn 1975, 1977). This should imply that the observed small eccentricity is very likely spurious.
However,  the system is probably found with a massive circumprimary disc remanent of a recent mass transfer burst and with the primary rotating at critical velocity (M12). It is possible that the observed eccentricity could be the result of a dynamical perturbation introduced by the circumprimary massive disc. This is an open question we let for a  future investigation.


\begin{table}
 \caption{Orbital elements for the donor of V\,393\,Scorpii obtained by minimization of
 the $\chi^{2}$ parameter given by Eq.\,(1) for the cases of circular (Case C) and elliptical (Case E) orbits .
 The value $\tau^{*} = \tau- 2454507.78$ is given and also the error measured as half the difference between the maximum and minimum quantity in one isophote
 1$\sigma$. }
 \begin{tabular}{@{}lccccccc@{}}
 \hline
\hline
 Case             &$P_{o}$ &$\tau^{*}$ &$e$  &$\omega$ &$K_{2}$ &$\gamma$  &$\chi^{2}$\\
              &(days) &    &     & &(\kms) &(\kms) & \\
\hline
C &7.7139&9.6&0.00 &1.95&170.7&-19.1&9 \\
E &7.7130 &1.3&0.04 &1.49&169.7&-18.6&2 \\
error &0.0020&0.1&0.02&0.08 &2.0 &2.8&-\\
\hline
\end{tabular}
\end{table}

\begin{figure}
\scalebox{1}[1]{\includegraphics[angle=0,width=8.5cm]{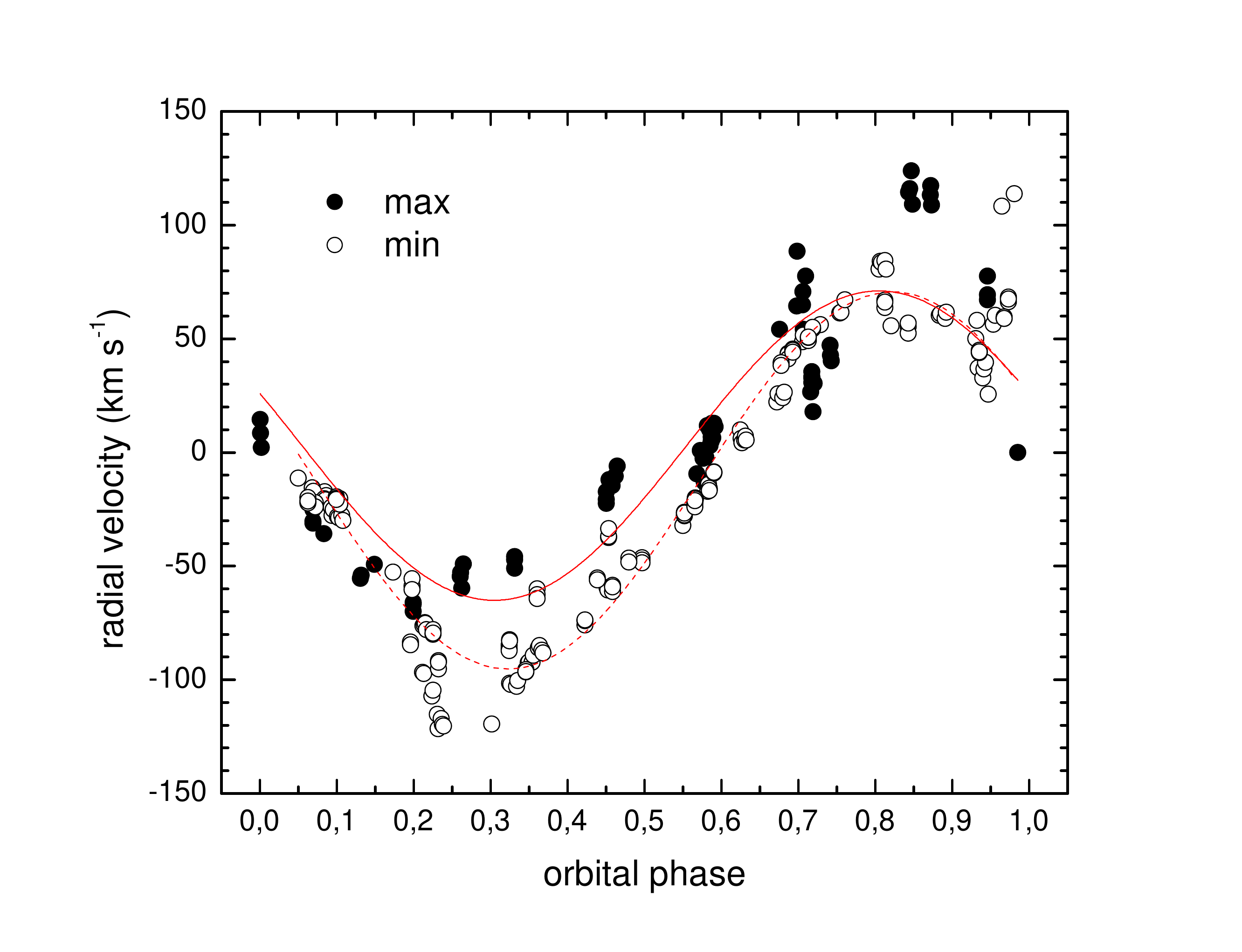}}
\caption{Up:  He\,I\,5875 radial velocities at different long cycle phases, 0.2 $< \Phi_{l} \leq $ 0.8 (circles) and 0.8 $< \Phi_{l} \leq $  0.2 (dots) and the best sinus fits. }
  \label{x}
\end{figure}

\subsection{The He\,I\,5875  line and the mass ratio}

Radial velocities for He\,I 5875  were measured using simple gaussian fits. The corresponding spectral region  is virtually uncontaminated by donor spectral features. The RVs are in anti-phase with the donor RVs, revealing their origin in or around the primary star (Fig.\,4).
There are large RV dispersion at a given phase, and the overall pattern of the RV curve changes during the long-cycle.  
This fact suggests that He\,I lines could be affected by residual emission at some epochs. It is therefore difficult to estimate the gainer radial velocity half-amplitude $K_{1}$ and the mass ratio from these RVs. We notice, however, that the half-amplitudes  are 82.9 $\pm$ 1.5  and 68.1 $\pm$ 4.5 \kms at low and high state, respectively (Table 6), much larger than $K_{1}$ = 45.6 $\pm$ 4.6 km s$^{-1}$ derived from the study of UV and IR lines by M10. If we assume that these half-amplitudes represent the motion of the more massive star then the larger values indicate a mass ratio $q \equiv M_{2}/M_{1}$  $\approx$ 0.4 or 0.5, whereas the M10 figure  indicates  $q$ = 0.27 $\pm$ 0.03. M10 preferred the lower value for the mass ratio since: (i) it allows the formation of a disc around the gainer and (ii) it is more compatible with the mass derived for the secondary star. We add a third reason: (iii) a low mass ratio value is compatible with synchronous rotation for the secondary star. In fact, for a secondary star corotating with the binary:\\

$\frac{v_{2r}sin\,i}{K_{2}} \approx (1+q)\frac{0.49q^{2/3}}{0.6q^{2/3}+\ln(1+q^{1/3})}$\hfill(5)\\

 
 \noindent
(Eggleton 2006, Eq. 3.9), where $K_{2}$ is the donor radial velocity half-amplitude. Using the above equation, $K_{2}$ =  170 \kms and $v_{2r} sin\,i$ = 60 km s$^{-1}$, we obtain $q \approx$ 0.27, consisting with the low $q$ found by M10. 

The strength and shape of the He\,I\,5875 line vary during the orbital and long cycles.  The  $FWHM$ increases systematically from \lp 0.5 to \lp 0.0 and decreases from high to low state. The $FWHM$ increases during the orbital cycle attaining a maximum at \op 0.9, similarly to He\,I\,10830  (M10).   The amplitude of the $FWHM$ variability is about 200 km s$^{-1}$. Emission at the flanks of the He\,5875 absorption line is seen occasionally,  with typical full peak separation $\Delta \lambda_{5875} \sim$  700 km s$^{-1}$. 

The long-cycle variability can be explained by filling emission increasing toward the high state.  
At low state we observe the deeper and narrower line, and at high state the line is almost filled by emission, producing the observed $FWHM$ increase.
 Accordingly, the equivalent width ($EW$) of this line anti-correlates with their $FWHM$. 
 The maximum at \op 0.9 is produced by an enhanced red absorption wing.
 This event coincides with the passage of the hot spot  across the line of sight. It could reflect large infall velocities in the absorbing material of the gas stream.






\subsection{Spectral disentangling and donor-subtracted spectra}

The  variable and multicomponent nature of the spectra make difficult their analysis and requires some kind of useful approximation to isolate single component contributions. In order to remove the donor light from the spectra we assume that its contribution to the total light is additive to the other light sources and represented by the model proposed by M12, i.e. it is independent of the long cycle. This assumption is justified because: (i) the amplitude of the long cycle is small compared with the large magnitude changes observed during the orbital cycle and (ii) we do not detect large systematic changes of the donor during the long cycle.
 
We constructed donor-subtracted spectra by removing the synthetic spectrum of the donor from the observed spectrum.  Firstly, the synthetic spectrum was Doppler corrected and scaled according to the  contribution of the secondary star at a given orbital phase and at the given wavelength range. For normalized fluxes:\\

\small
$ f_{res} (\lambda, \Phi_{o}, \Phi_{l}) = f (\lambda, \Phi_{o},  \Phi_{l}) - p(\Phi_{o}, \lambda_{c}) \times f_{d} (\lambda, \Phi_{o}) $\hfill(6)\\
\normalsize

\noindent
where $f_{res}$ is the donor-subtracted flux, $f$ the observed flux, f$_{d}$ the  synthetic donor spectrum, $p$ is the 
 fractional contribution  of the donor  derived from our model and $\lambda_{c}$ the representative wavelength where this factor is calculated. Theoretical $p$-factors correctly account for the variable  projected area of the donor and its larger flux contribution at longer wavelengths (Fig.\,5). The method described above effectively removed the main contribution of the donor to the observed spectra. The result was a set of  ``donor-subtracted'' spectra that were normalized to the new continuum. 

\begin{figure}
\scalebox{1}[1]{\includegraphics[angle=0,width=8cm]{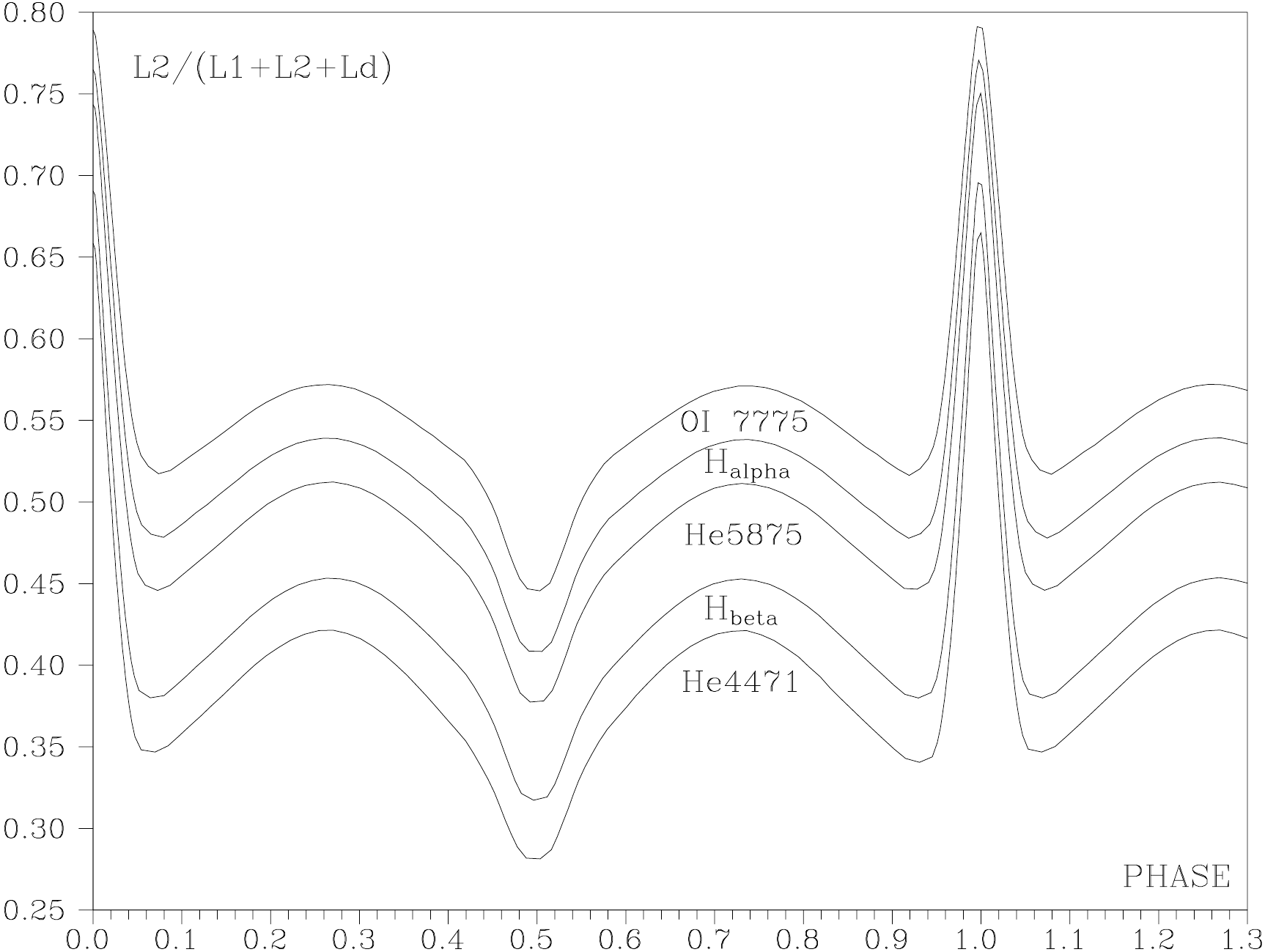}}
\caption{Light contribution factor for the donor at different spectral lines and orbital phases according to the M12 model. $L_{1}$, $L_{2}$ and $L_{d}$ are the gainer, donor and disc fluxes, respectively.}  
  \label{x}
\end{figure}

 
Disentangling of the gainer and the  disc was not intended since there are no model available for the disc spectral distribution and  probably its spectrum combine in a complex way with the spectrum of the primary.

\subsection{The He\,I\,4471 and Mg\,II 4482 lines }

The ratio  between the equivalent width  of He\,I\,4471 and Mg\,II\,4482  is a good temperature indicator in early B-type stellar atmospheres, being larger for late B-type stars. We calculated equivalent widths and equivalent width ratios ($R$) as a function of stellar effective temperature for our grid of synthetic stellar spectra (Fig.\,6). These calculations are useful to get some insight on the physical conditions 
of the region forming the helium absorption.

$EW_{4471}$  and $R$ show orbital and long-cycle variability.
During high state magnesium emission increases faster, 
producing a larger $R$ at this state (Fig.\,7). This result suggests the $R$ could be used as a rough diagnostic for the long cycle stage in absence of a clear ephemeris,  for old spectroscopic 
observations for instance. The orbital variability of $R$ and $EW_{4471}$ at low state suggests that the effective temperature for the hot component varies with the orbital phase and it is larger at
\op 0.4.
At low state $EW_{4471}$  and $R$ are  compatible with line formation in a photosphere with $T_{hot}$ = 16.000 $K$. This is the same temperature found for the inner disc region (and the gainer) by M12. In contrast, at high state $R$ is larger than any theoretical value, mostly due to the weakness of the magnesium line by filling emission. This makes uncertain any $R$-based temperature   at this stage.

\begin{figure}
\scalebox{1}[1]{\includegraphics[angle=0,width=8.5cm]{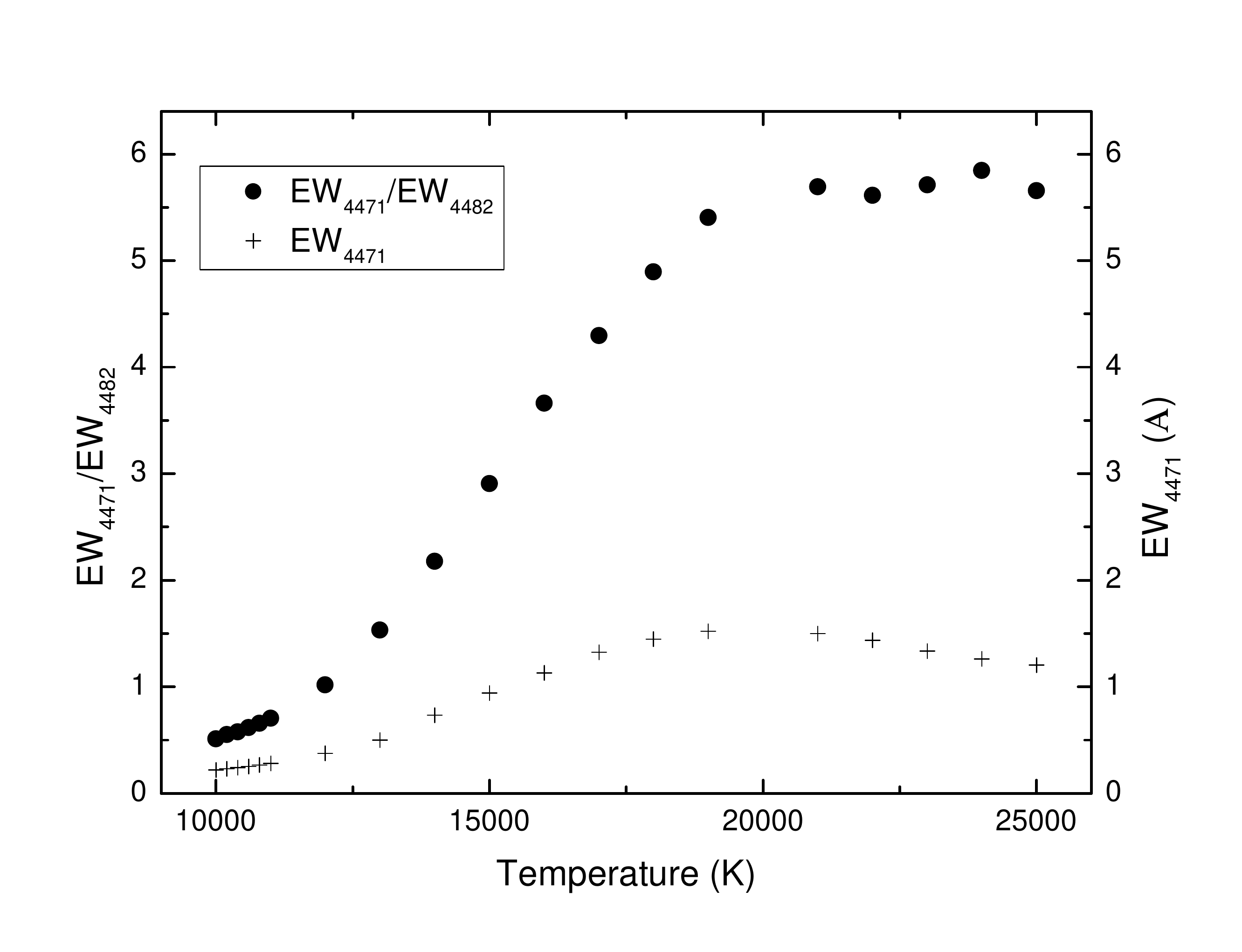}}
\caption{Equivalent  widths and equivalent width ratios for synthetic spectra calculated for stellar atmospheres with $\log g$  = 4.0, $V_{turb}$= 2 km s$^{-1}$ and solar metallicity.   }
  \label{x}
\end{figure}


\begin{figure}
\scalebox{1}[1]{\includegraphics[angle=0,width=8.5cm]{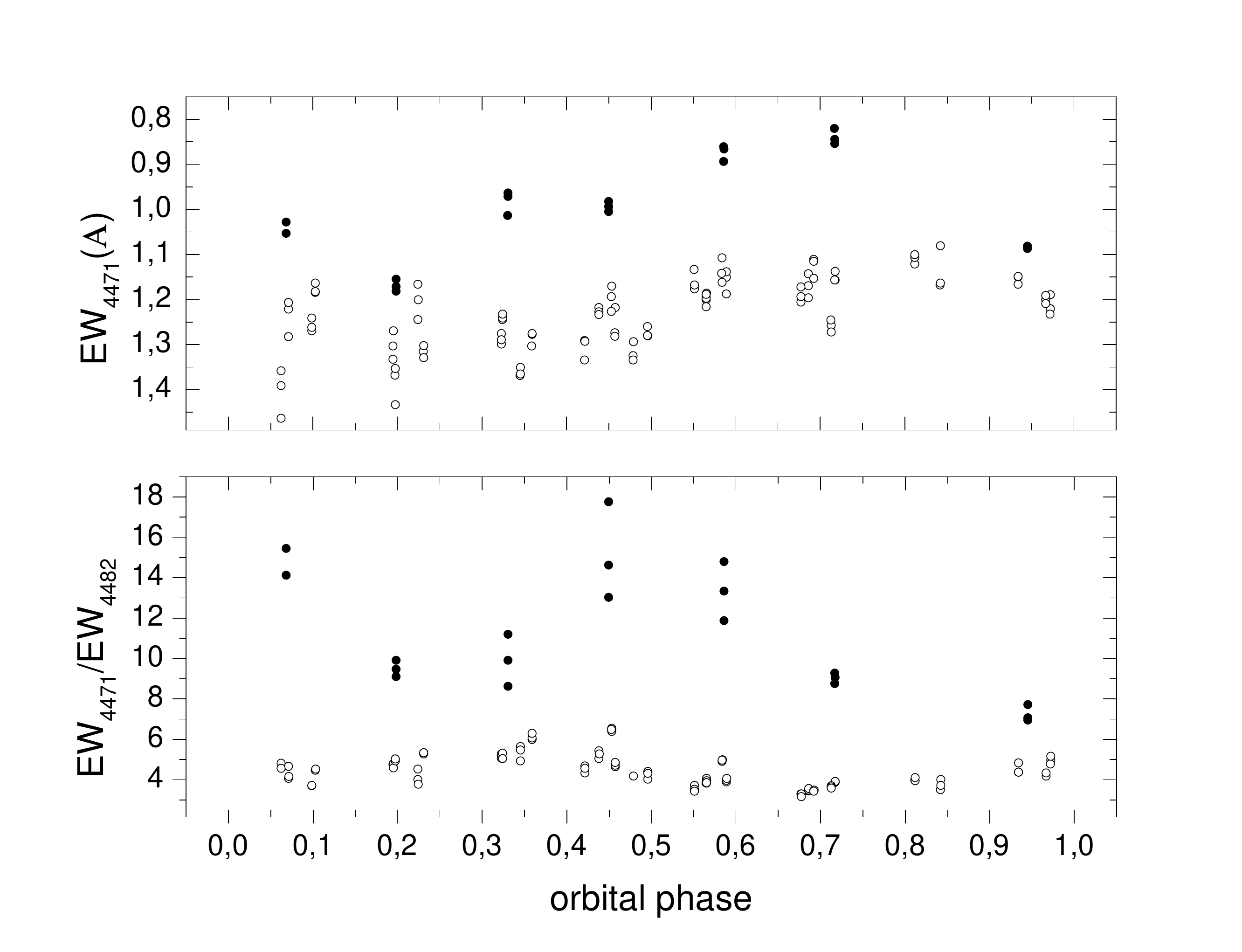}}
\caption{He\,4471 equivalent width and He\,4471/Mg\,II\,4482 equivalent width ratio versus orbital phase. UVES data for 0.8 $<$ $\Phi_{l}$ $<$ 0.2 (filled circles) and 0.2 $\leq$ $\Phi_{l}$  $\leq$ 0.8 (open circles) are shown.}
  \label{x}
\end{figure}

\begin{figure}
\scalebox{1}[1]{\includegraphics[angle=0,width=8cm]{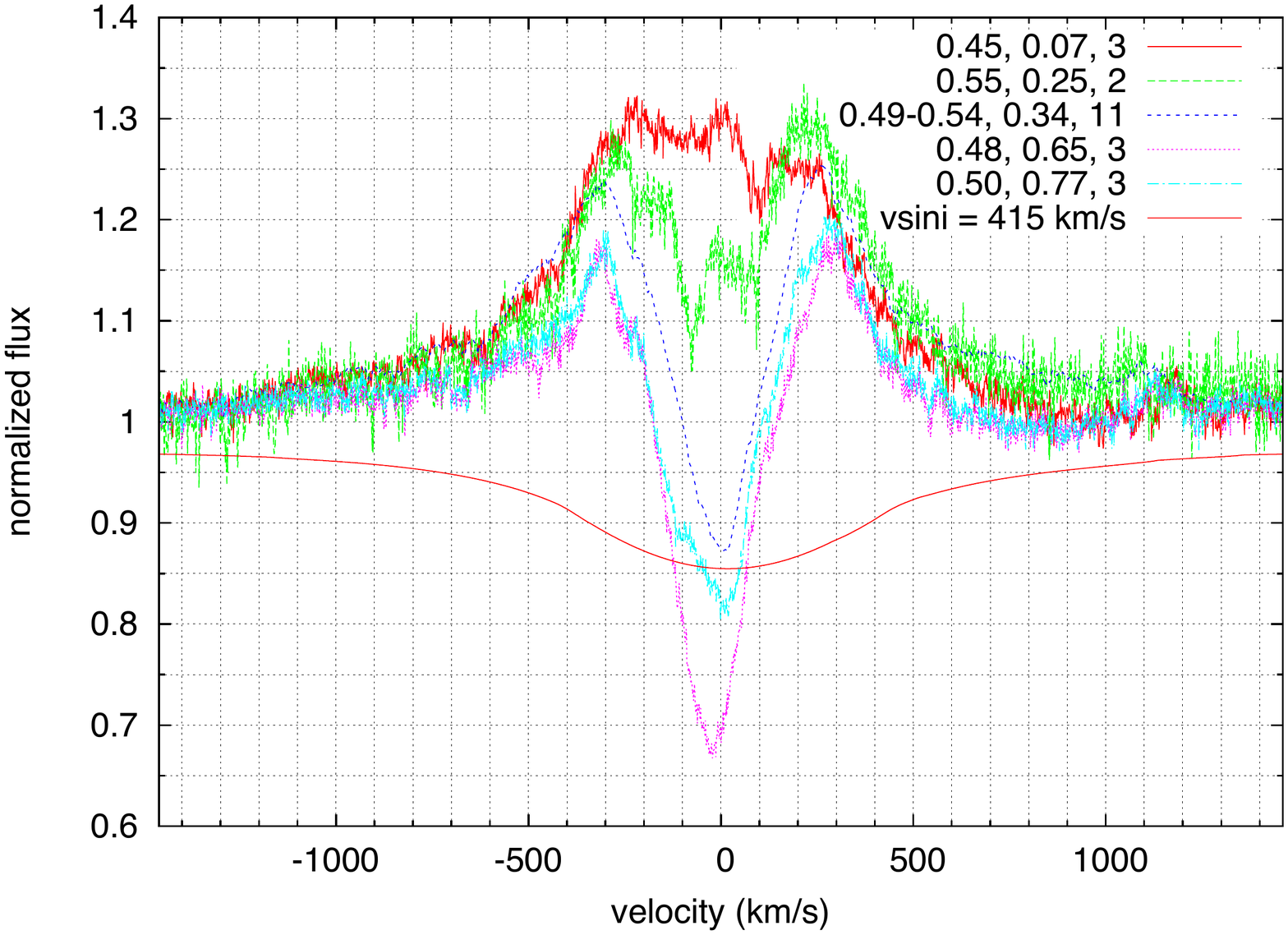}}
\scalebox{1}[1]{\includegraphics[angle=0,width=9cm]{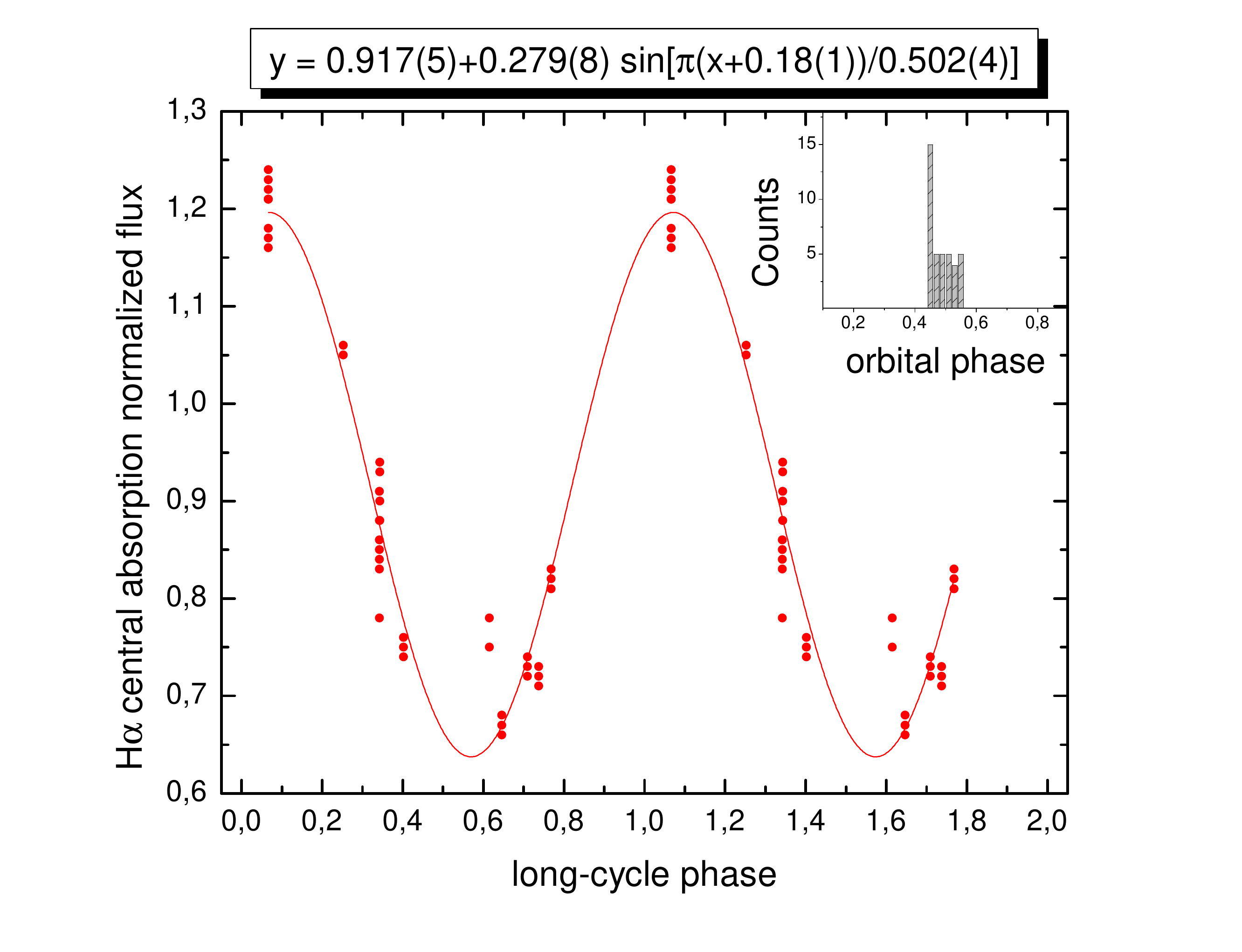}}
\scalebox{1}[1]{\includegraphics[angle=0,width=8cm]{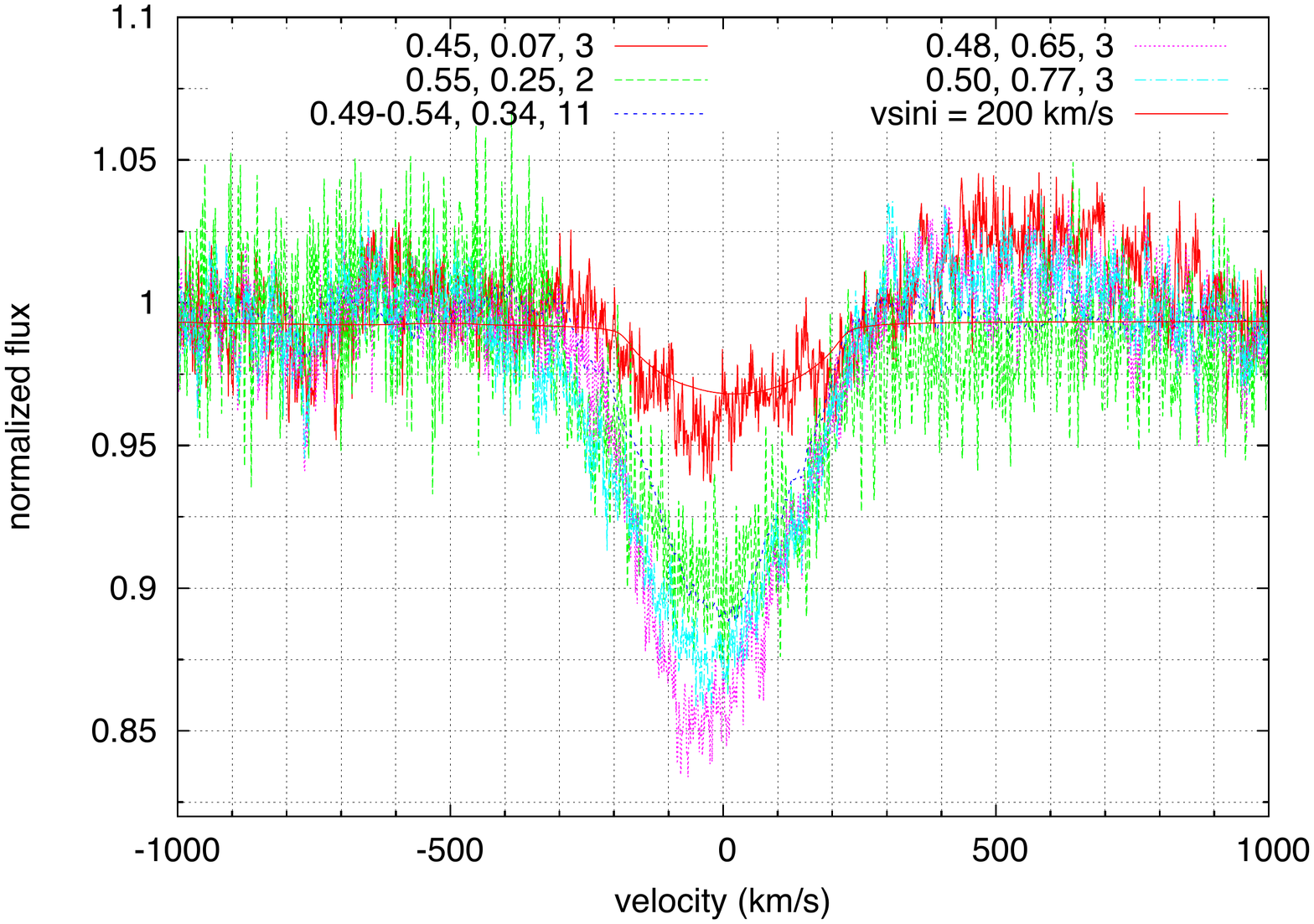}}
\caption{Up and down: Donor-subtracted H$\alpha$ and He\,I 6678 profiles near  secondary eclipse at different  long cycle phases. Labels indicate, from left to right,  orbital phase, long cycle phase and number of spectra averaged. Synthetic profiles  of the gainer  calculated with parameters given by M12  are shown for different rotational velocities. 
X-axis velocities are with respect to the system center of mass.  Middle: The behavior of the intensity of the central absorption of the donor-subtracted H$\alpha$ profile at secondary eclipse during  the long cycle. The best sinus fit is also shown along with the histogram of orbital phases for the considered  spectra. }
  \label{x}
\end{figure}

\subsection{Balmer and He\,I\,6678 lines}

\subsubsection{General considerations}

According to the M12 model,  at  \op 0.5 occurs the largest occultation of the donor by the disc, that attains  maximum visibility.  We compare H$\alpha$ and He\,I\,6678 \AA~profiles at this phase 
 for  several  long-cycle epochs and conclude that (Fig.\,8): 
(i) at low state H$\alpha$ shows a relatively narrow and deep central absorption (CA) flanked by emission peaks of similar intensity. CA is up to 65\% the continuum level at this orbital phase, (ii) when the system goes to high state the global emission  strength increases and the CA tends to disappear. After a careful examination of the whole dataset, we concluded that this behavior produces the variability of the peak separation described later, i.e., the stronger the CA, the larger the peak separation. Some flat-toped profiles observed at long maximum fit this tendency.

We compare  theoretical photospheric spectra for a critically rotating gainer with the observed H$\alpha$ and He\,I\,6678 central absorption profiles and conclude that 
they do not match at long cycle minimum (Fig.\,8). However, at high state the He\,I  profile fits the theoretical  profile of a slower gainer.  We propose that this result is only the  consequence of the evident filling emission. In this view the broadening of 200 \kms probably reflects the kinematics of the gainer pseudo-photosphere. Others interesting features observed in these spectra are the  H$\alpha$ emission wings extending beyond 1000 \kms (see Section  4.2).

The general variability for these lines is illustrated in Fig.\,9. The 
helium line is filled by emission at high state, a characteristic also seen in other helium lines. It is notable the presence of emission flanks  with $\Delta \lambda_{6678} \sim$ 900 \kms.  
These flanks are also seen with varying intensity in other helium lines. 

The H$\alpha$ line appears as a deep absorption surrounded by weak emission shoulders during low state but it is a strong and double emission (or in few occasions triple or irregular)  during high state. For double emission lines the violet emission peak changes its relative intensity with respect to the red peak during the orbital cycle. This property is measured through the $V/R$ ratio, where $V$ is  the violet peak intensity and $R$ the red peak intensity. The $V > R$ condition near \op 0.75  reveals a larger H$\alpha$ emitting region at the binary hemisphere containing the hot spot (Fig.\,9). Note also in Fig.\,9 the presence of a weak emission line redward H$\alpha$ that we identified with C\,I\,6588. This line follows the donor RV curve and will be discussed in Section 3.9.2.

The H$\beta$ and H$\gamma$ emission profiles show similar behavior to H$\alpha$, but with weaker emission and most importantly with very  extended emission wings (EEWs), beyond $\pm$ 2000 km s$^{-1}$ (Fig.\,10).  These  velocities are too large to be explained in terms of Keplerian rotation around the gainer. The EEWs usually appear at the red profile flank during the first half of the orbital cycle and at the blue flank  during the 2nd half, consisting with an emitting region in the binary hemisphere containing the hot spot, a region high enough to be observed as an emitting source rather than an absorbing media. As occurs with the low-velocity Balmer line emission, the strength of these EEWs is larger at high state. 

Occasionally we observe a narrow emission feature redward He\,I\,4921   that is sometimes accompanied with a redshifted absorption (Fig.\,10). We identify this feature as C\,I\,4932; its RV suggests an origin in or around the secondary star. This line  will be discussed   along with other weak spectroscopic features in a forthcoming paper.



\begin{table}
\centering
 \caption{Results of the sinusoidal fits ($\gamma + K \sin(2\pi(x-\delta)$)) to the central absorption RV curves. The root mean square of the fit is also given. The parameters $\gamma$,
 $K$ and $rms$ are given in \kms.}
 \begin{tabular}{@{}lrccc@{}}
 \hline
Line (state) &$\gamma$  &$K$  & $\delta$ &$rms$\\
\hline
H$\alpha$  (all)  & -7.9 $\pm$ 1.5   &59.6 $\pm$ 2.3  &0.550 $\pm$ 0.005 &20.1\\
H$\beta$  (low)  &   -9.8 $\pm$ 1.0    &64.0 $\pm$  1.5& 0.546 $\pm$ 0.004&15.1 \\
H$\beta$    (high)& -4.1 $\pm$  3.5     &78.4 $\pm$ 5.4  & 0.525 $\pm$ 0.009&32.1\\
H$\gamma$ (low) &-15.8 $\pm$ 1.1      & 71.3 $\pm$  1.5   &  0.529 $\pm$ 0.003&14.7\\
H$\gamma$ (high) & -14.8 $\pm$ 2.7  &83.2 $\pm$ 4.4  &  0.517 $\pm$ 0.006&26.6\\
O\,I\,7773 (low) &-13.5 $\pm$ 2.6 & 32.1 $\pm$ 3.8 & 0.593 $\pm$ 0.017 &20.7\\
He\,I\,5875 (low) &-12.3  $\pm$ 1.1  & 82.9 $\pm$ 1.6  & 0.572 $\pm$ 0.003&15.3 \\
He\,I\,5875 (high) &3.0  $\pm$ 3.1  & 68.1 $\pm$ 4.5  & 0.555 $\pm$ 0.010&25.0\\
He\,I\,6678 (all) &  -8.8 $\pm$ 1.3  &70.2 $\pm$ 1.9  &0.554 $\pm$ 0.004& 21.6\\
\hline
\end{tabular}
\end{table}

\subsubsection{Radial velocities of emission line features}

Here we present the analysis of the radial velocities of the central cores and emission peaks of the Balmer lines and some helium lines.
The CA RV curves were fitted with sinus functions with the best fitting parameters given  in Table 6. In general, the radial velocities for Balmer and helium features 
are in anti-phase with the donor RVs,  indicating an origin somewhere near the gainer (Figs.\,11 and 12). However, they do not follow the expected motion of the gainer. The half-amplitudes are usually larger than the adopted $K_{1}$ value
and the positions of the light-centers, as measured by the $\delta$ parameters,  are displaced from the  center of the primary. We notice larger Balmer amplitudes during high state and the opposite for Helium lines. We also notice the relatively small amplitude of the O\,I\,7773 line.  The line centers, as measured by the $\gamma$ parameter, have larger velocity than the systemic velocity, which can be interpreted in terms of an outflow or wind. 

Contrary to He\,5875, the He\,I\,6678 CA shows small changes through the long cycle. The RV of the blue emission peak shows a quasi-sinusoidal behavior with large amplitude, whereas the red peak   remains almost stationary at 390 \kms shifting  to 550 \kms  at high state. On low state the  blue H$\alpha$ emission  peak shows a large blueshift at \op 0.25, as occurs with He\,I\,5875 (Fig.\,4). This  
might be due to the approach of the bright spot at longitude 162$^{\circ}$ (M12). Another notable feature is the shift 
to bluer wavelengths of  the H$\alpha$ CA around \op 0.15,  as corroborated by data obtained on high state at 3 different observatories.  This anomaly  is also observed in the asymmetric absorption wings discussed in Section  4.5. 

If we use the system inclination $i = 80^{\circ}$  (M12), and the basic kinematics formula:\\

$\frac{r}{R_{\odot}} = \frac{Pv}{50.633} $ \hfill(7)\\

\noindent
where $r$ represents the radial distance from the center of rotation, $P$ the rotational period in days and $v$ the linear orbital velocity of material moving in the orbital plane measured in \kms, then we
can find the position of the light-centers of the lines listed in Table\,6. 
We find all line light centers located practically in the line joining the stellar centers. In addition, light centers for Balmer and helium lines are located between 9 and 13 $R_{\odot}$ from the barycenter measured in the direction of the gainer;  for O\,I\,7773 this figure is 4.9 $R_{\odot}$.  Remembering that the center of mass is located at 7.16 $R_{\odot}$  from the gainer and that this star has a radius of 4.4 $R_{\odot}$ (M12), these results show that all lines are formed somewhere around the gainer,  O\,I\,7773 probably near the gainer hemisphere facing the donor (consistent with the location of the DACs forming region) and the other lines in more distant regions.  Since the gas dynamics is very likely not limited to the orbital plane and includes important vertical motions not considered in this analysis we should take these results with caution.


\subsubsection{Equivalent widths and peak separations}

On low state the equivalent width of the  He\,I\,6678 line shows large orbital variability and increasing emission during the main eclipse (see Fig.\,11, similar behavior is observed in He\,I\,4471 in Fig.\,7). This simple observation reveals that the source of the line  emission is not eclipsed as  occurs with the continuum flux of the donor-subtracted spectra.   Interestingly, the same parameter shows much weaker orbital variability on high state, and importantly the strong variability during eclipse disappears. This behavior is difficult to understand, but it could be partly explained if the extra continuum light present at the high state is not eclipsed. In other words, the wind not only contributes to line emission at low/high state but also to continuum emission at high state, i.e. it is optically thicker during the long maximum.

As mentioned before, the H$\alpha$ peak separation is smaller at high state, principally due to the weaker CA. Moreover, the equivalent width of  Balmer lines vary during the long cycle in anti-correlation with the peak separation (Fig.\,13). The H$\alpha$ peak separation  goes from 690 at low state to 340 \kms at high state, where it cannot be due to Keplerian motions since  it should imply an orbit outside the gainer Roche lobe.
For H$\beta$ and H$\gamma$ the corresponding figures are  800 to 500 \kms and  1000 to 650 \kms, respectively. This behavior is opposite to the observed in He\,6678, where the peak separation increases on high state (Fig.\,10). 


If the CA is formed in the gainer disc-like extended photosphere, then we can reasonably assume that rotational broadening determines the CA width, and also indirectly the peak separation.
The maximum rotational velocity in the disc is derived from $\Delta \lambda$$_{\gamma}$ at low state (about 500 \kms) and suggests 
that the primary is in fact rotating at critical velocity.
As for helium lines, the Balmer emission  increases during main eclipse,  indicating that the main source for Balmer line emission is not eclipsed neither. We measure
maximum equivalent widths at main eclipse of  -40, -20 and -20 \AA, for H$\alpha$, H$\beta$ and H$\gamma$, respectively (Fig.\,14). 



\begin{figure}
\scalebox{1}[1]{\includegraphics[angle=0,width=8.5cm]{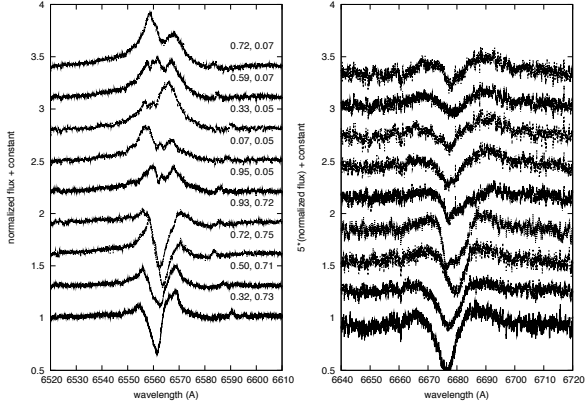}}
\caption{Donor-subtracted H$\alpha$ (left) and He\,I\,6678  (right) spectra at selected epochs. Orbital phases (left) and long-cycle phases  (right) are given. Note the C\,I\,6588 emission redward H$\alpha$.}
  \label{x}
\end{figure}

\begin{figure}
\scalebox{1}[1]{\includegraphics[angle=0,width=8cm]{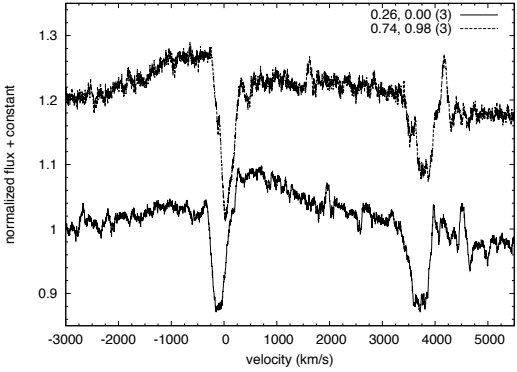}}
\caption{Donor subtracted spectra showing H$\beta$  extended emission wings  up to $\pm$ 2500 km s$^{-1}$.  The He\,I\,4921 absorption line is also visible along with narrow C\,I\,4932 emission. The numbers between parenthesis indicate the number of averaged spectra. }
  \label{x}
\end{figure}

 \begin{figure}
\scalebox{1}[1]{\includegraphics[angle=0,width=8cm]{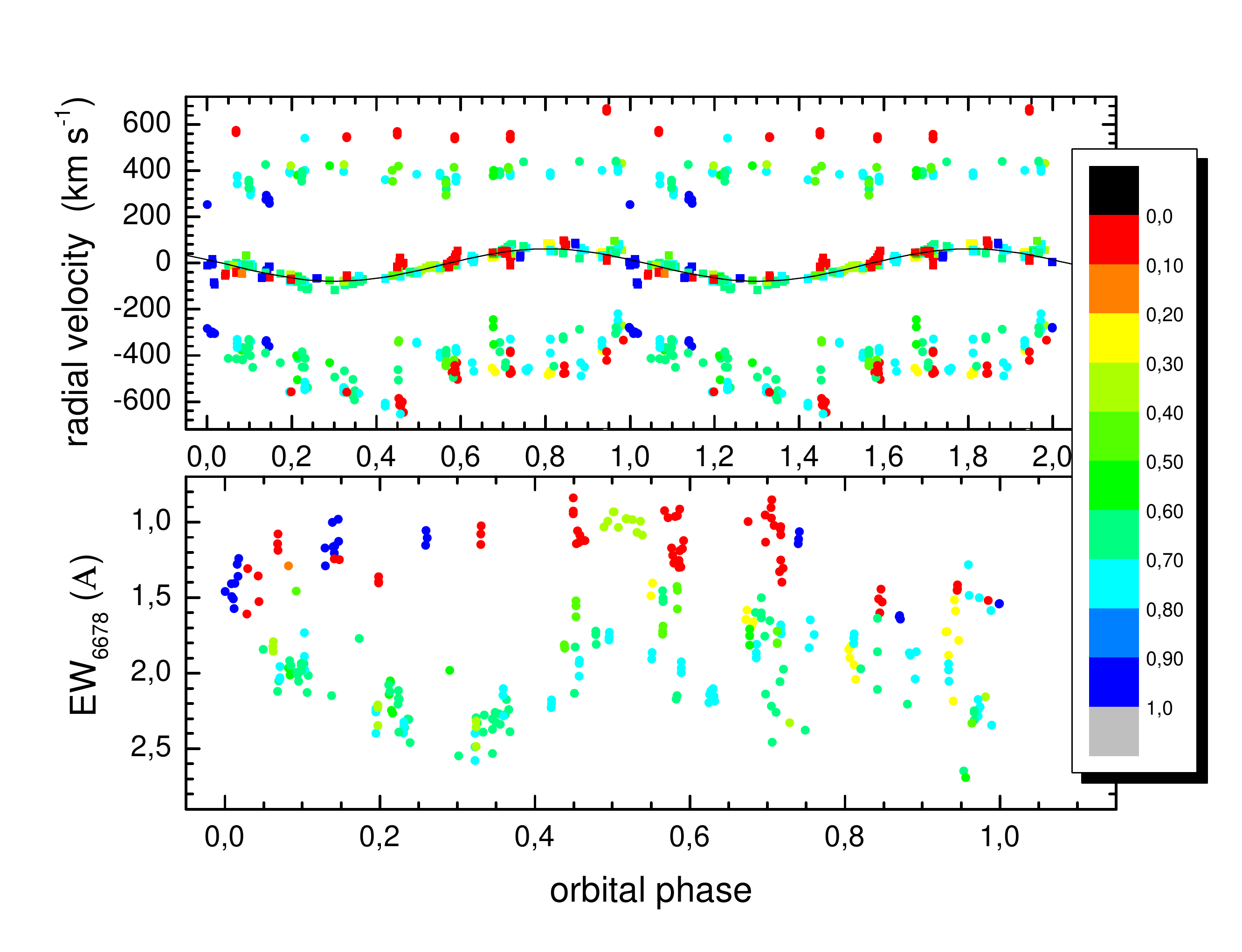}}
\caption{Up: Radial velocity for the central absorption and the red and blue peaks of the He\,I\,6678 line. The best sinus fit for the central absorption is also shown. Down:  The equivalent width for the He\,I\,6678 as a function of the orbital and long cycle phase. The color  band represents $\Phi_{l}$ ranges.}
  \label{x}
\end{figure}

\begin{figure}
\scalebox{1}[1]{\includegraphics[angle=0,width=9cm]{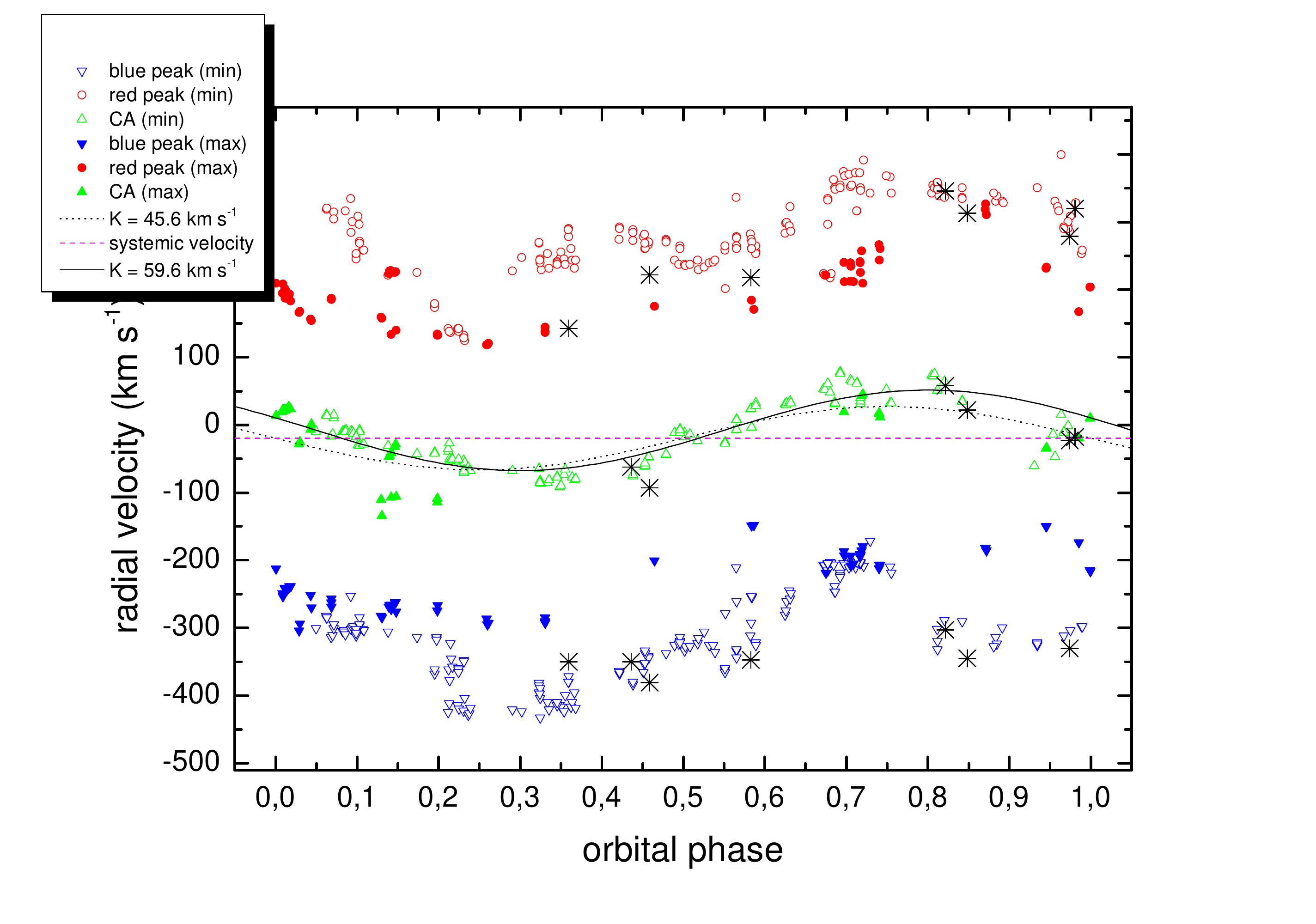}}
\scalebox{1}[1]{\includegraphics[angle=0,width=9cm]{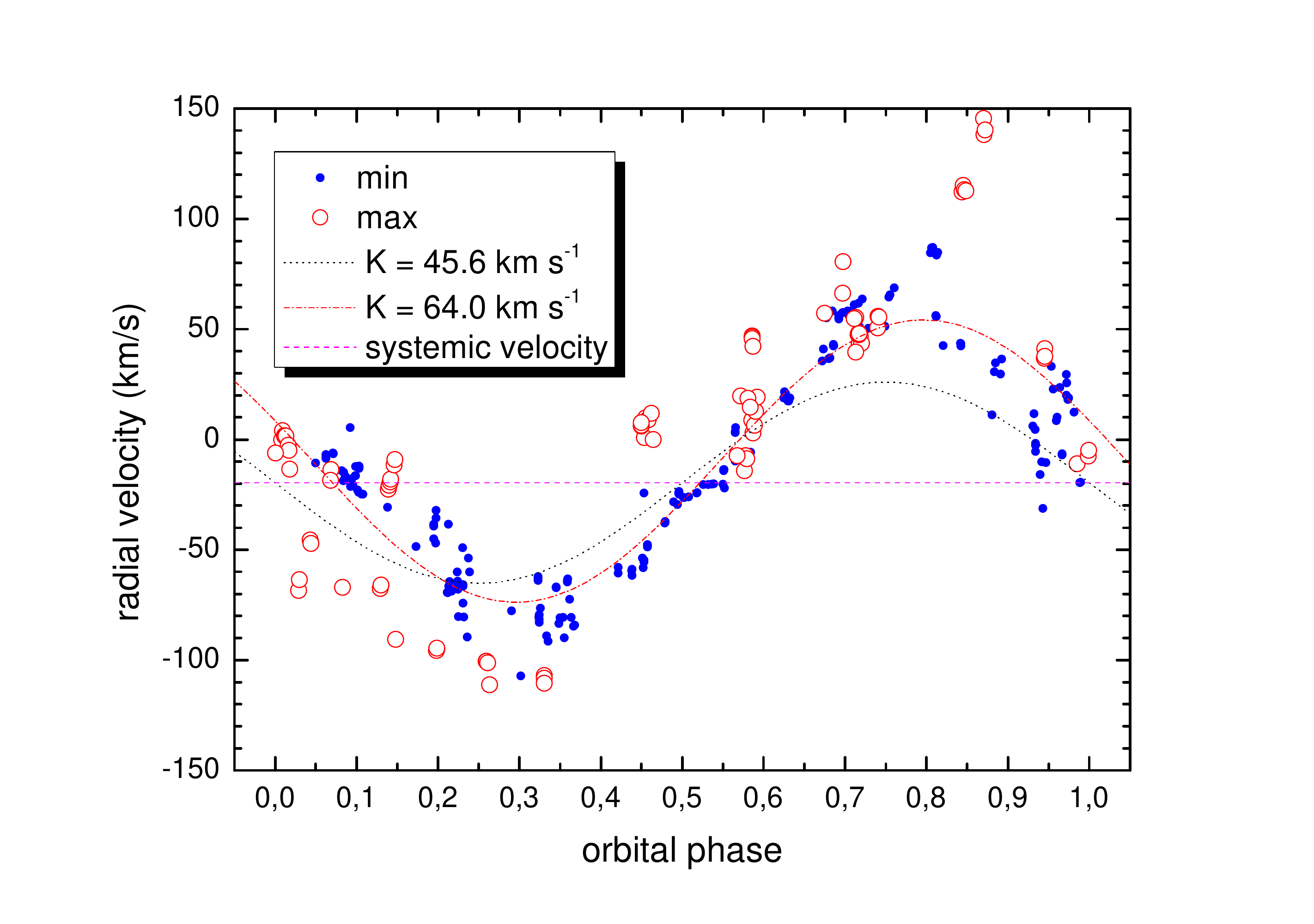}}
\scalebox{1}[1]{\includegraphics[angle=0,width=9cm]{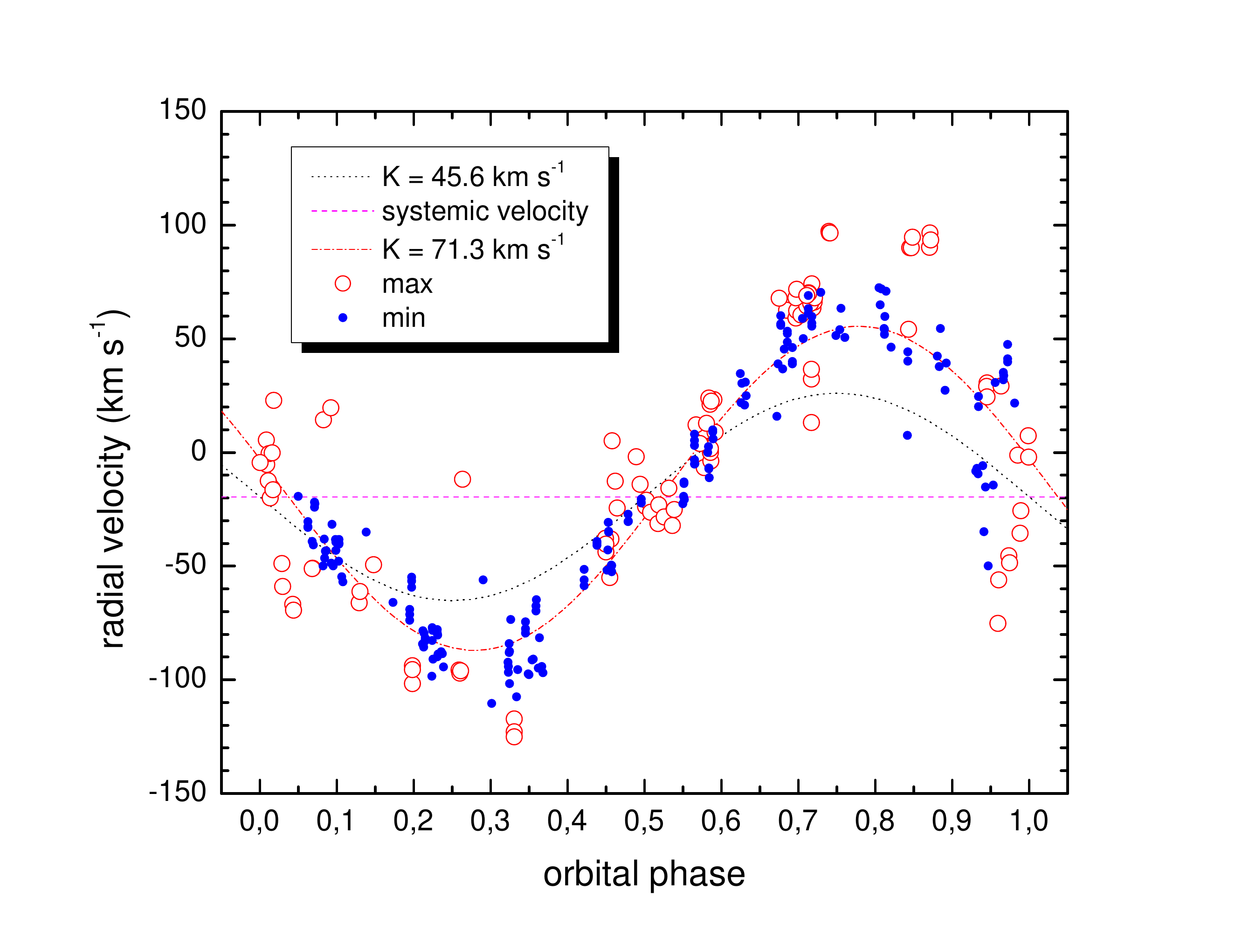}}
\caption{Up: Radial velocity for the blue and red peak of the H$\alpha$ emission and the central absorption. The best sinus fit (excluding outliers around \op 0.15) along with the assumed track for the gainer are also shown. KPNO data  are shown as asterisks and are not included in the fit. Middle and Down: The radial velocity for the H$\beta$ (middle) and H$\gamma$ (down) central absorption. The best sinus fits on low state are shown, along with the assumed track for the gainer.  In all graph data for 0.8 $<$ $\Phi_{l}$ $<$ 0.2 (max) and 0.2 $\leq$ $\Phi_{l}$ $\leq$ 0.8 (min) are shown with separated symbols. }
  \label{x}
\end{figure}

\begin{figure}
\scalebox{1}[1]{\includegraphics[angle=0,width=8cm]{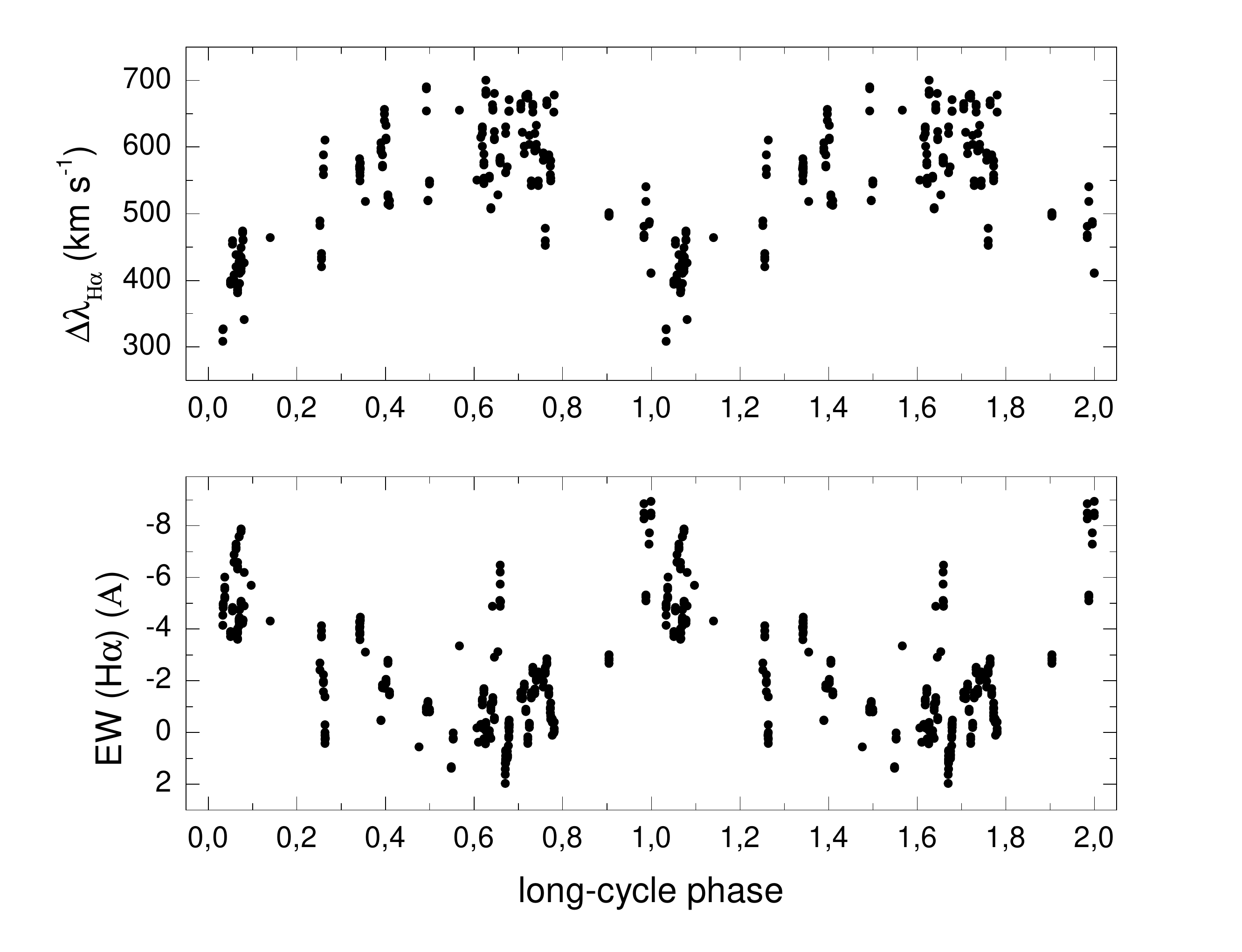}}
\caption{The peak separation (up) and  equivalent width (down) of the donor-subtracted  H$\alpha$ line profile.  
Data for 0.95 $<$ $\Phi_{o}$ $<$ 1.05 are not shown.  }
  \label{x}
\end{figure}

\begin{figure}
\scalebox{1}[1]{\includegraphics[angle=0,width=8cm]{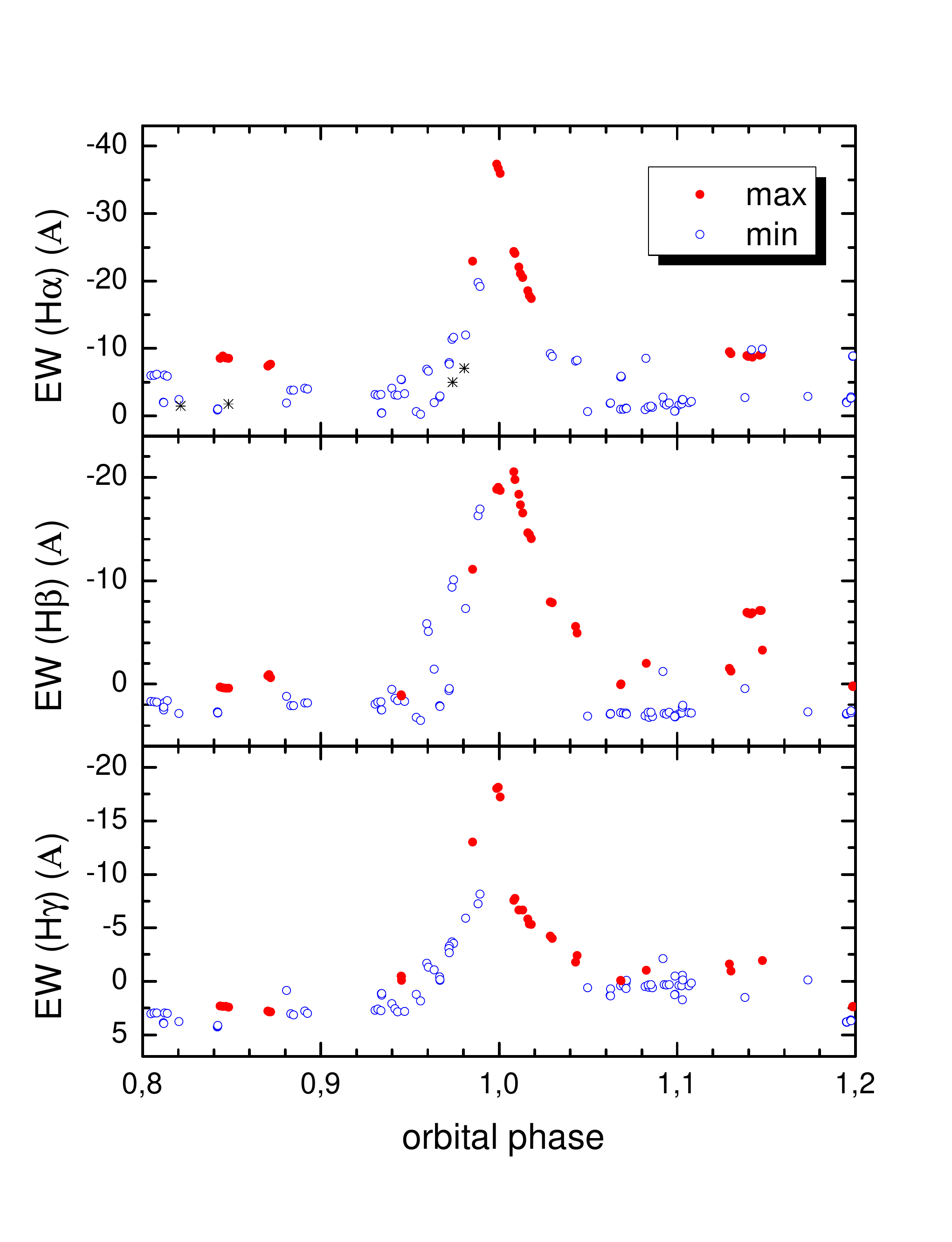}}
\caption{The equivalent width of the donor-subtracted  Balmer line profiles during main eclipse. Data on the long cycle phases 0.8 $<$ $\Phi_{l}$ $<$ 0.2 are shown as filled circles and those with 0.2 $\leq$ $\Phi_{l}$ $\leq$ 0.8 as open circles. Asterisks indicate KPNO data.}
  \label{x}
\end{figure}

\begin{figure}
\scalebox{1}[1]{\includegraphics[angle=0,width=9cm]{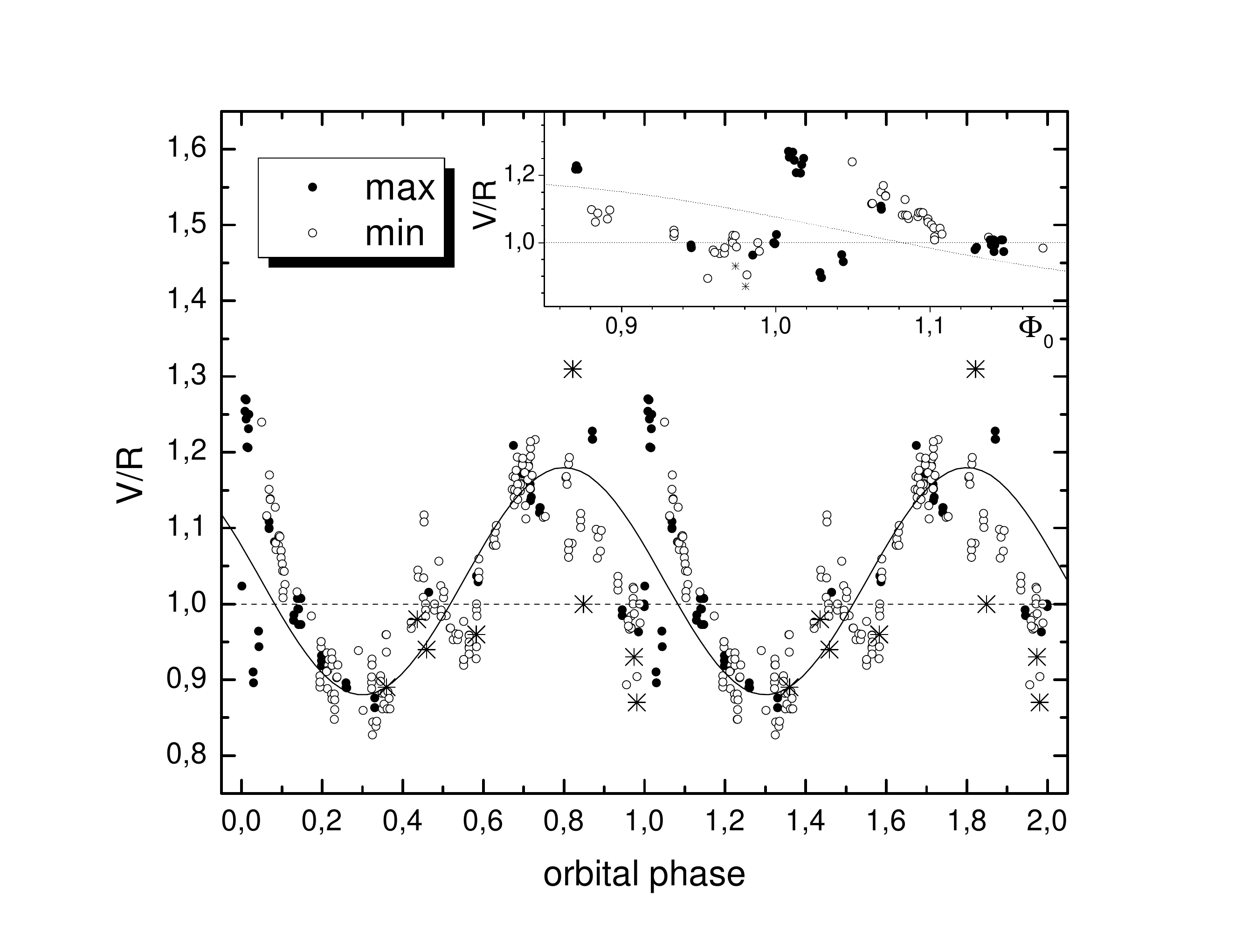}}
\caption{The ratio between the violet and red peak intensity  of the donor-subtracted H$\alpha$ double emission profiles (the inset shows 
a zoom around the main eclipse).  The sinusoid $V/R$ = 1.03 - 0.15 sin(2$\pi$($\Phi_{o}$-0.05)) has been plotted as reference. Data with 0.8 $<$ $\Phi_{l}$ $<$ 0.2 (max) are shown as filled circles and those with 0.2 $\leq$ $\Phi_{l}$ $\leq$ 0.8 (min) with open circles. Asterisks indicate KPNO data. }
  \label{x}
\end{figure}

 
  
  \subsubsection{H$\alpha$ V/R variability}
  
The H$\alpha$ $V/R$ ratio follows a quasi-sinusoidal pattern  during the orbital cycle with local maxima at \op 0.05, 0.45 and 0.70 (Fig.\,15).  In the same figure a reference sinusoid  represents the smoothed  orbital $V/R$ variability. This variability can be explained by the displacement of the CA across a stationary emission profile. This view is  difficult to test measuring   the  supposedly  stationary profile extension, due to their  strength variability and gradual merging with the continuum, but it is supported by the fact that the $V < R $ branch coincides with the  blue-shifted CA position and the $V > R$ one with the red-shifted.  The stability of the $V/R$ curve  at high and low state reflects the stability of the disc where the CA is formed, a result also supported by our light-curve photometric model (M12). 

The zero point of the V/R curve is larger than unity, indicating the predominance of the $V >  R$ condition. This is consistent with an outflow of the emitting region, as also indicated by the $\gamma$ values
in Table\,6. The $V/R$ curve shows 3 enigmatic successive and rapid  reversals just after main eclipse. In addition, 
a ``s-type'' variation is found around the referential sinusoid between \op 0.7-1.2 and a ``z-type'' perturbation between \op 0.4-0.7. 

 In order to interpret  the s-type  and z-type  waves, 
 we assume that the high latitude region responsible for Balmer emission extends to the disc of such way that the lower parts merge with the disc bringing its rotational velocity. In this case the s-wave can be viewed as the eclipse of the approaching disc edge by the donor during \op 0.7-1.0 and the eclipse of the receding edge during \op 0.0-0.2. In the same picture the z-wave probably shows the eclipse {\it by the disc} of the receding (approaching) hot spot region in \op 0.4-0.5 (0.5-0.6). We remember that according to the light curve model a hot spot is expected in the place where the gas stream hits the disc. This  hot spot faces the observer at \op 0.9  (M12) and probably produces, as already mentioned, the $FWHM_{5875}$ peak at \op 0.9.

In summary, our observations seems to indicate that a wind is emerging from the disc producing Balmer emission lines by electronic recombination.  We notice that while our wind-disc model explains well the Balmer line variability, the  alternative hypothesis of a
circumbinary disc cannot  explain many of the observed features.


\begin{figure}
\scalebox{1}[1]{\includegraphics[angle=0,width=8cm]{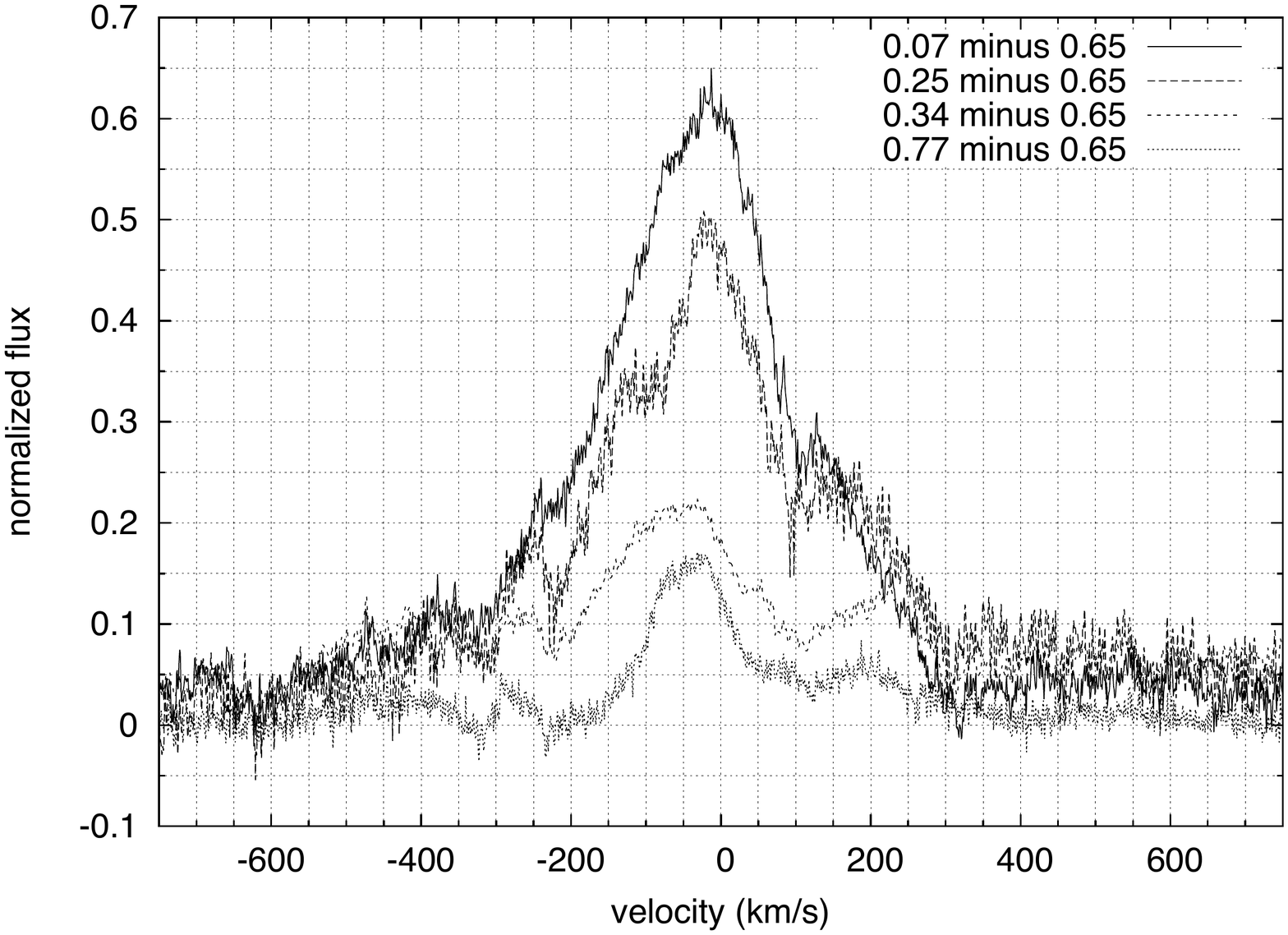}}
\scalebox{1}[1]{\includegraphics[angle=0,width=8cm]{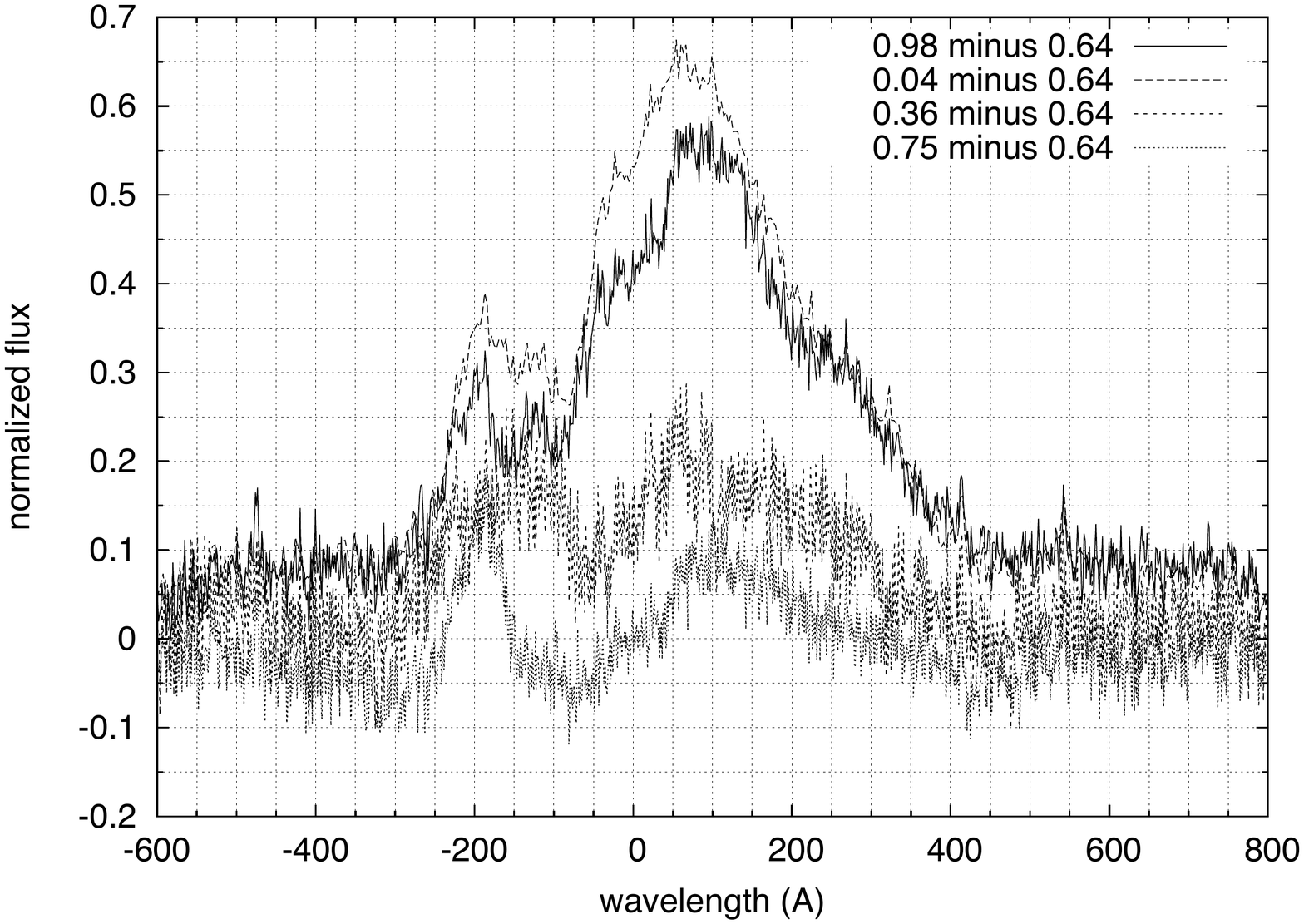}}
\caption{Up: H$\alpha$ difference spectra near $\Phi_{o}$= 0.00 at selected long cycle phases. Down: H$\alpha$ difference spectra near $\Phi_{o}$= 0.75 at selected long cycle phases. Velocities are with respect to the system center of mass. }
  \label{x}
\end{figure}



\subsection{Balmer difference spectra}

In order to analyze the variable line emission we constructed {\it difference spectra}
by subtracting profiles from a reference profile obtained roughly at the same orbital phase but different long cycle phase, usually  that one with the minor line emissivity.
We do not correct for the underlying variability of the continuum through the long cycle, since it should have a minor effect on line profile intensity. 
Difference spectra constitute useful representations of line emission changes at a given orbital phase and projected velocity bin
during the long cycle. The results for H$\alpha$ discussed here are basically the same that for  H$\beta$ and H$\gamma$.

All difference spectra for H$\alpha$ taken around \op 0.0 indicate that the emission grows  principally at low projected velocities when the system approaches the high state (Fig.\, 16). 
The characteristic single peak profile of the extra emission does not correspond with disc emission, either circumprimary or circumbinary and reinforces the thesis of a wind.


At \op 0.75 the emission grows around  -200 km s$^{-1}$ and 50 km s$^{-1}$ (Fig.\,16).
These velocities suggest that emitting material exist around the donor and the primary, a fact corroborated by the H$\alpha$ Doppler map later. 
In general, Balmer difference spectra show that the wind line emissivity increases on the high state.

\subsection{Doppler Tomography}

Doppler tomography was introduced as a  tool for the study of accretion discs by Marsh \& Horne (1988). This analysis method is now a widespread procedure in the study of emission lines in Algols (Richards 2004), and provides a quantitative mapping of optically thin line forming regions in the velocity space. When severe self-absorption and/or intrinsic line broadening are present then the tomograms provide at least a concise and convenient way of displaying phase-resolved line profile measurements. The Doppler reconstructions presented in this paper were computed using the filtered back-projection method (Rosenfeld \& Kak 1982). 
The two- dimensional reconstruction of structures in the orbital plane has a unique solution that depends only on the two-dimensional data set (i.e. $f_{v}$ ($\Phi_{o}$, $\lambda$) where $f_{v}$ is the flux).
  Our code has been successfully applied in a number of cases (e.g. Mennickent, Diaz \& Arenas 1999) and the comparison of observed and reconstructed line profiles shows the self-consistency of the method.

 When interpreting Doppler maps we should keep in mind the difficulty in the representation of optically thick structures and components  above or below the orbital plane.
This is specially relevant for the Balmer emission lines discussed in this paper. The evidence presented in this section for H$\alpha$ is not conclusive, and the map should be only taken as a particular two-dimensional representation of the three-dimensional mass flows in the binary.

\subsubsection{The H$\alpha$ Doppler map}
We built the Doppler map for the donor-subtracted  H$\alpha$ line with resolution $FWHM$ = 65 km s$^{-1}$. Only spectra with 0.8 $ < \Phi_{l} < 0.3$ were used, since others showed too weak emission. The map is shown in Fig.\,17. The stream path plotted in the map is for a point mass primary. For a 4.4 $R_{\sun}$ primary radius the stream would ultimately collide with the star at tomogram velocities
$V_{x}$ =  -675 km s$^{-1}$ and $V_{y}$  = -310 km s$^{-1}$  (i.e. in the lower left quadrant) or even sooner with disc or wind.

The H$\alpha$  absorption profile goes below the continuum
during some phases. This is inconsistent with an optically thin emission formed in the gas velocity field.
Having said that 3 interpretations for the Doppler map remain: (i)  the gas is optically thick in the lines, (ii)
the photospheric spectrum has a broad absorption component and (iii)  a combination of both is present (an optically thick  wind and/or disc).
The map indicates that the
absorption has a $FWZI$ of circa 150 km s$^{-1}$ and has significant
radial velocity with phase still consistent with the primary. 
We computed phased spectrograms and the orbital
velocity modulation of the absorption can be confirmed 
at high state.  It is possible to interpret this region as the optically thick circumprimary  disc. Additional pieces of information are that the emission almost disappears at $L_{1}$ and around \op 0.65. Finally, emission is detected around the donor, confirming the results inferred from our H$\alpha$ difference spectra. 
In summary, and taking into account our complementary spectroscopic analysis, the H$\alpha$ Doppler map is consistent  (but not conclusive) with an optically thick circumprimary disc plus high-latitude wind and emitting material located around the donor star.

\subsubsection{The  Mg\,II\,4481 and C\,I\,6588 Doppler maps}

We detected Mg\,II\,4481 and C\,I\,6588 emissions in the donor subtracted spectra whose velocities indicated an origin around the secondary star. These emission are very variable in intensity
and sometimes show double peaks. They are present at high and low state. A renormalization of the continuum was made around these lines and Doppler maps were constructed in the usual way. 

The Doppler map for Mg\,II\,4482 was centered at 4481.2 \AA, it shows that this line is  formed in the donor and it is not produced by  irradiation of the hemisphere facing the gainer, but probably corresponds to chromospheric activity (Fig.\,18). The same result is obtained with the C\,I\,6588 line. The presence of near/underlying absorptions 
in this line and the data sampling and spectral noise produces strong streaks mimicking emission at $L2$, $L3$ and the gas stream region. These are very likely artifacts of the reduction process.

The discovery of  chromospheric  emissions open the possibility that the donor in \var posses a non-neglected magnetic field.

\begin{figure}
\scalebox{1}[1]{\includegraphics[angle=0,width=9cm]{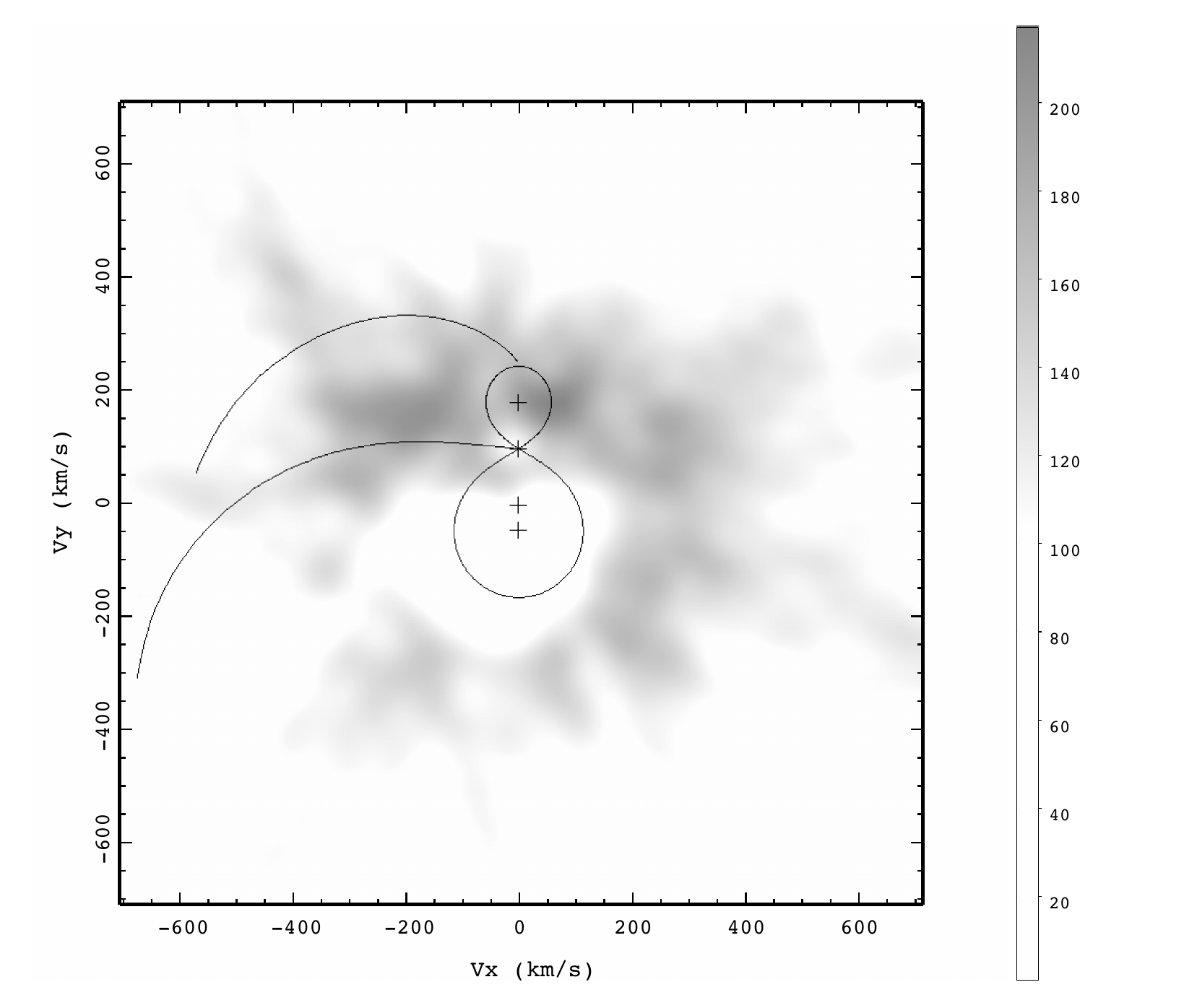}}
\caption{The back-projection Doppler map for the donor-subtracted   H$\alpha$ line during high state  (0.8 $< $ $\Phi_{l} $ $<$ 0.3). Crosses show, from up to down, donor center of mass, L1 point, system center of mass and gainer center of gravity. Roche lobe surfaces for a point mass gainer are shown along with the gas stream path  (lower track). The upper track  represents
the keplerian velocity of a disc if it would exist along the stream.}
  \label{x}
\end{figure}

\begin{figure}
\scalebox{1}[1]{\includegraphics[angle=0,width=9cm]{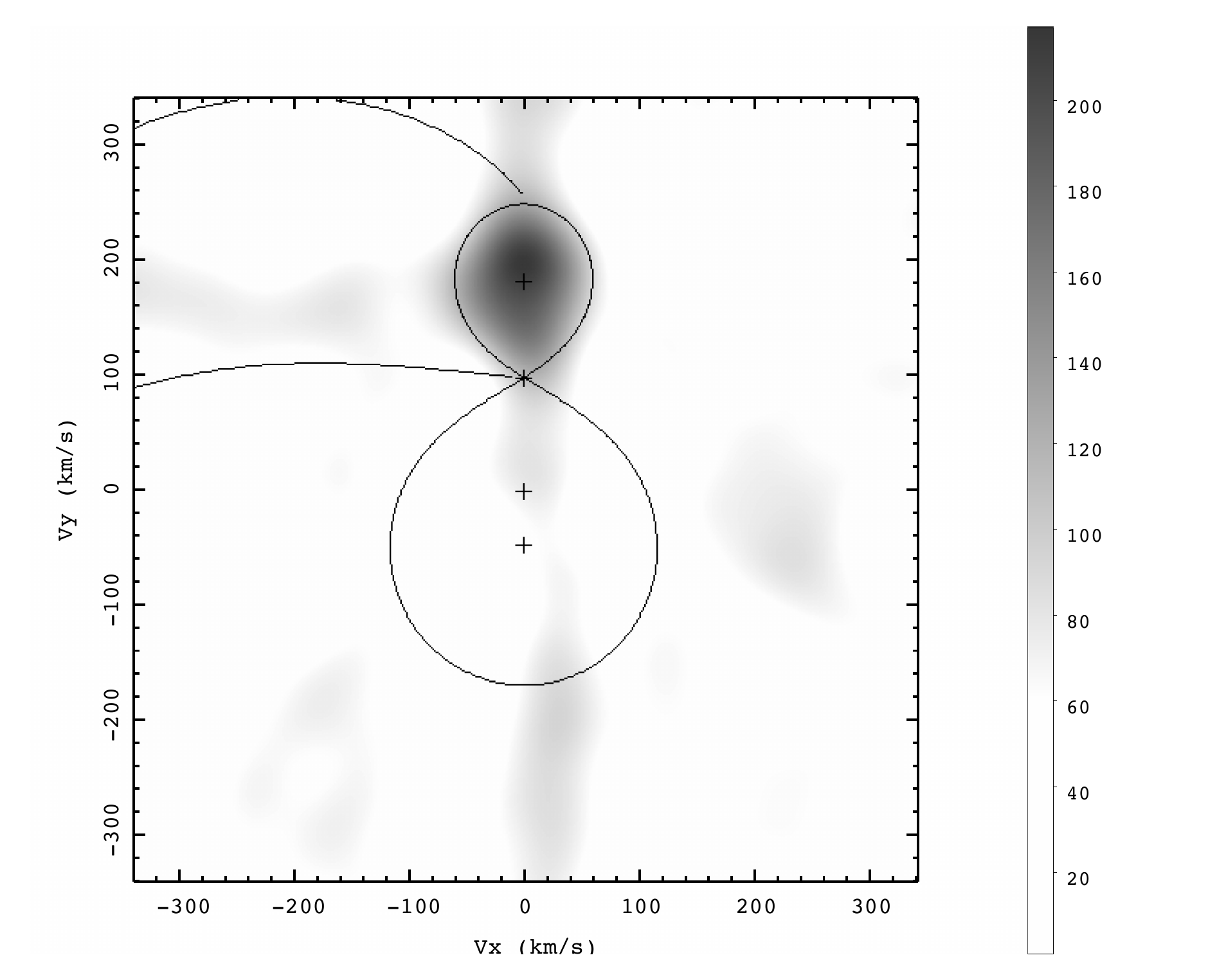}}
\scalebox{1}[1]{\includegraphics[angle=0,width=9cm]{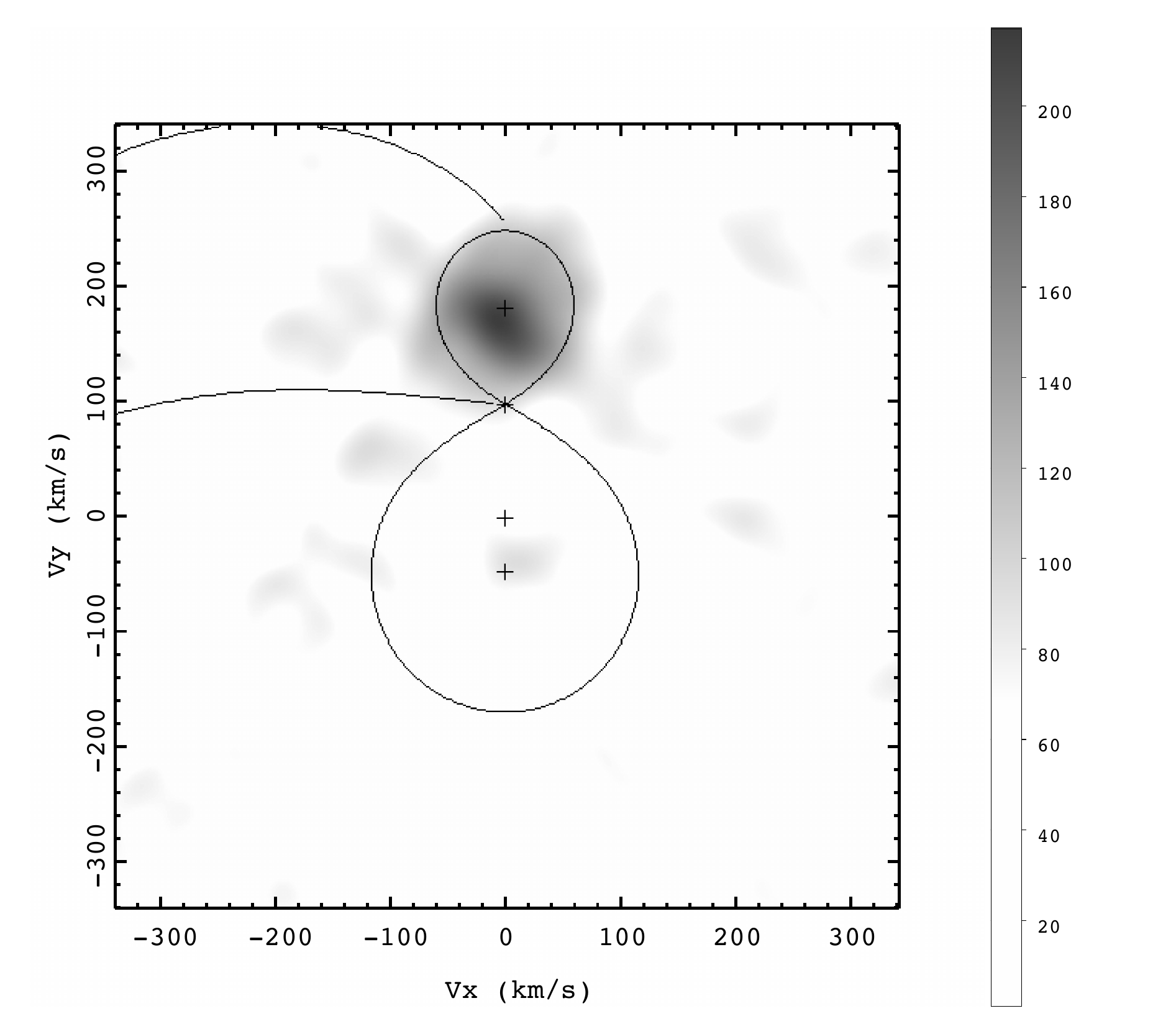}}
\caption{Up: The back-projection Doppler map for the donor-subtracted   Mg\,II\,4482 line. Symbols are as in previous figure. The $FWHM$ resolution is 60 \kms. Down: Same as above but for the C\,I\,6588   line with resolution 40 \kms.}
  \label{x}
\end{figure}

\subsection{The O\,I\,7773 triplet and the Si\,II  6347--6371 doublet}

The O\,I\,7773 line appears in absorption on low state and almost completely filled by emission on high state.  This line is a triplet with the redder component displaced by 67 \kms and 107 \kms from the two other components, but in our spectra it appears broadened and the single components are not resolved.
The line shows pronounced {\it blue} absorption  wings in the phase range \op 0.55--0.84 and enhanced {\it red}  absorption wings during \op 0.20--0.45 (Fig.\,19). These line depressions are similar to those observed in He\,I\,10830 (M10)  and are present even before applying the disentangling process.

 We detected  Discrete Absorption Components (DACs, of typical depth 3 to 5 \% of the continuum intensity) in virtually all O\,I\,7773 
 spectra with enhanced absorption  wings,  but in much larger number in the blue wing.
 DACs are visible even before spectral disentangling  and in spectra free of fringes and bad pixel regions.
From the above,  and from their visibility at specific orbital phases, we argue that DACs are not artifacts of the disentangling process or instrumental setup.





\begin{figure}
\scalebox{1}[1]{\includegraphics[angle=0,width=8cm]{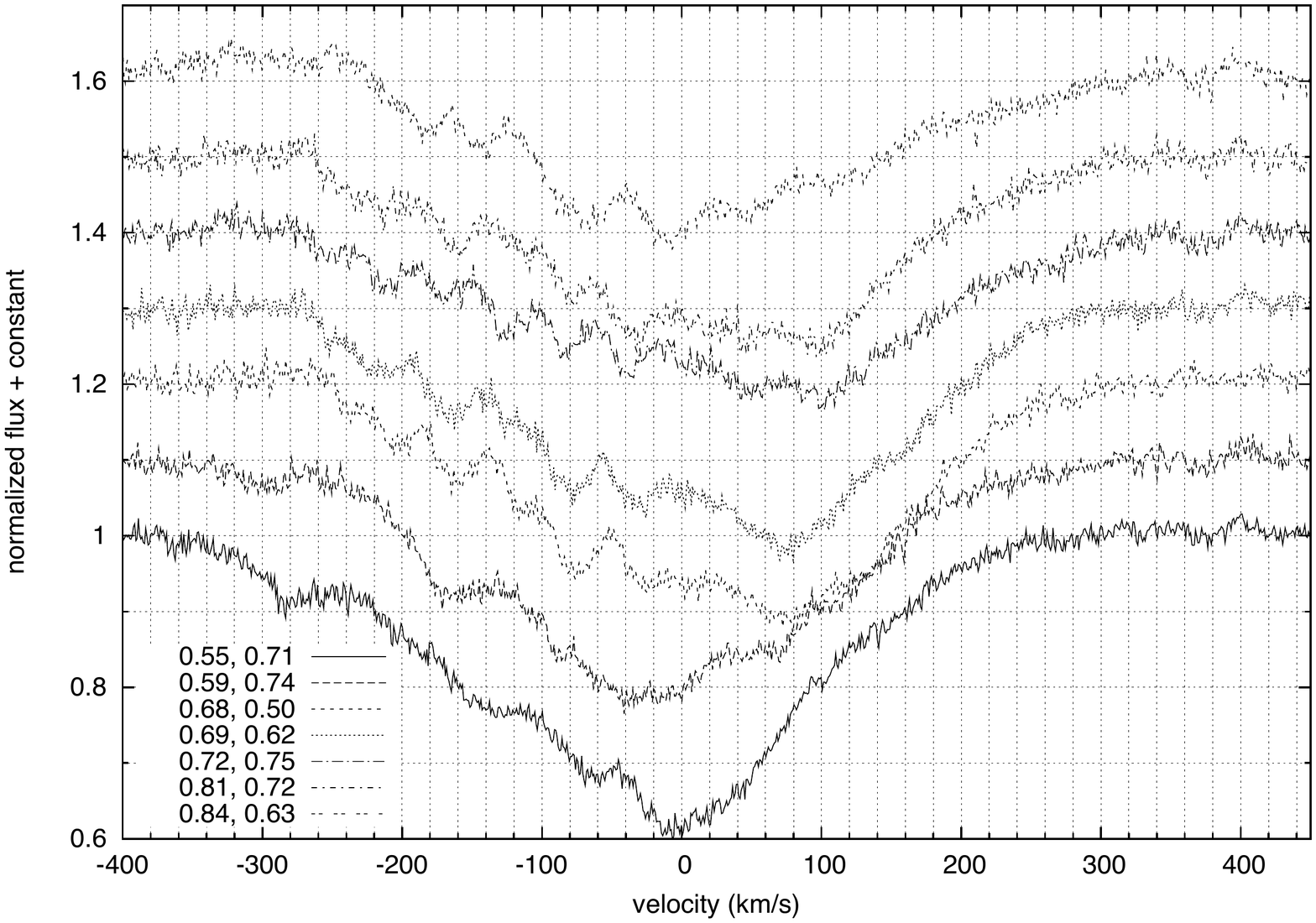}}
\scalebox{1}[1]{\includegraphics[angle=0,width=8cm]{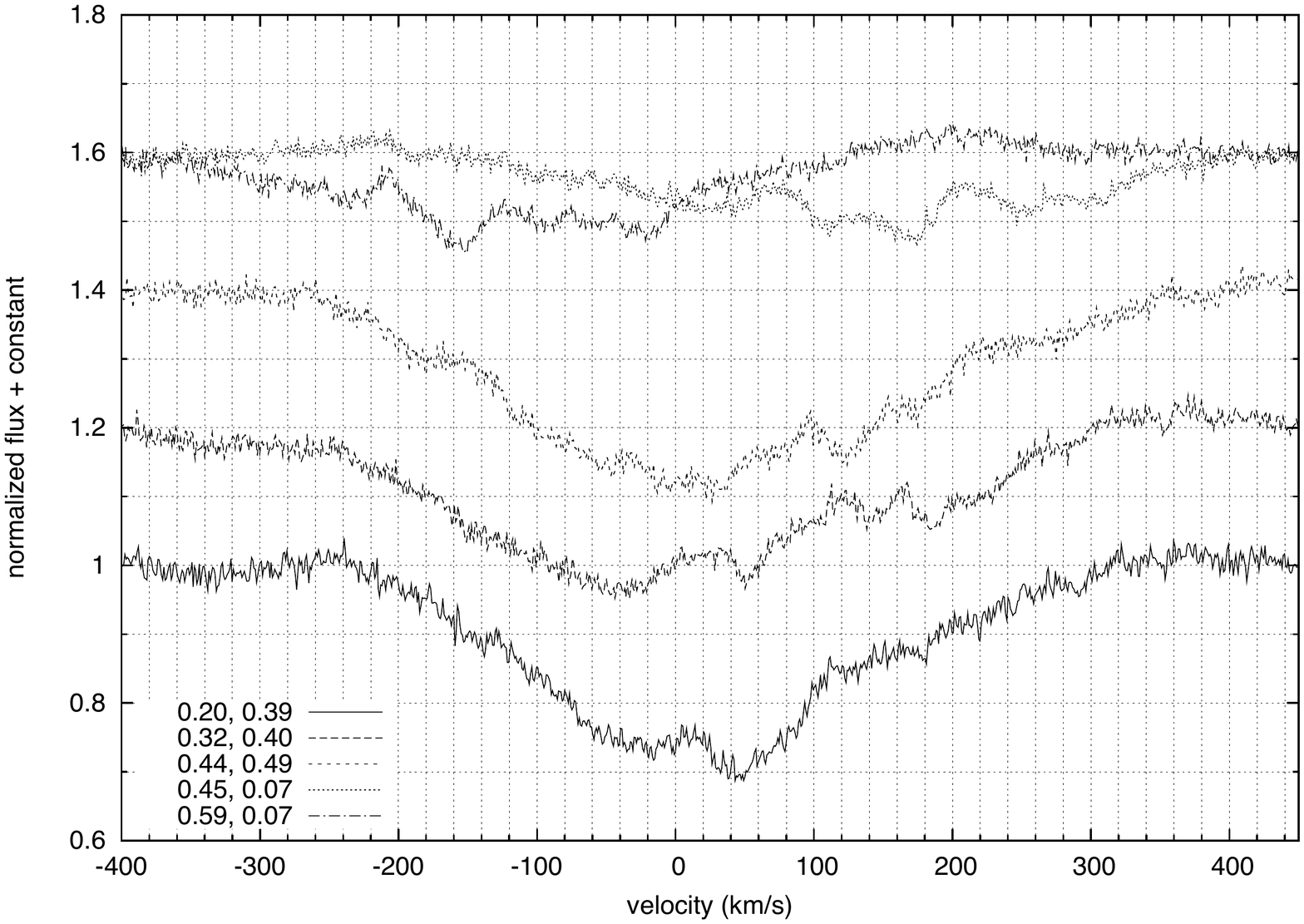}}
\scalebox{1}[1]{\includegraphics[angle=0,width=8cm]{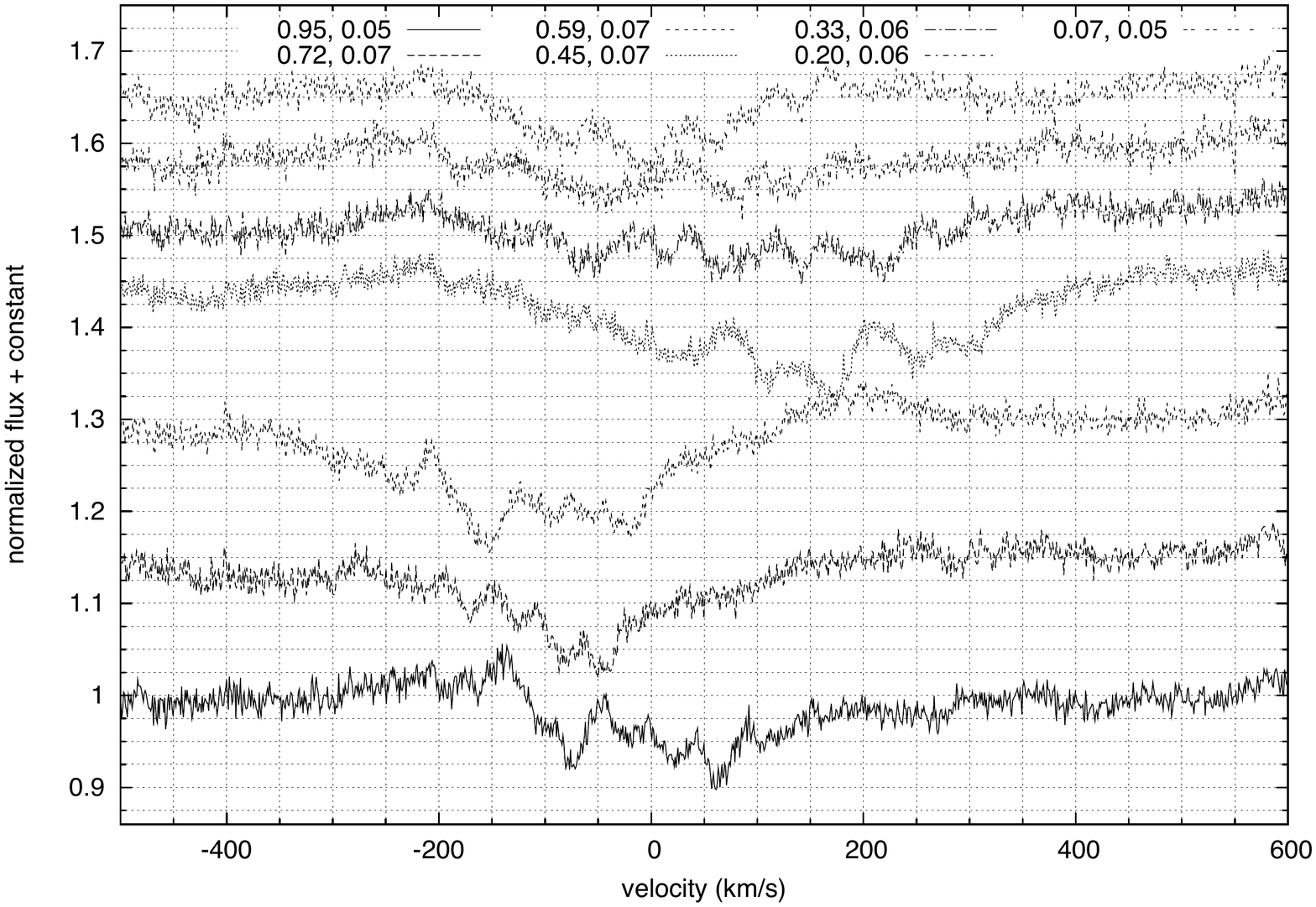}}
\caption{Up: Examples of donor-subtracted OI\,7773 lines showing DACs in enhanced blue wing. Middle: Examples of donor-subtracted OI\,7773 lines showing DACs in enhanced red wing. Down: Donor-subtracted OI\,7773  spectra during the high state. Labels indicate orbital and long-cycle phases.}
  \label{x}
\end{figure}

We averaged consecutive spectra taken few minute apart on a same night in order to improve the signal to noise ratio (typically $S/N \approx$ 120).  Then we measured DAC radial velocities  with respect to the reference wavelength   $\lambda_{0}$ = 7773.83 \AA.
Up to 8 DACs were detected in a single spectrum but more could be  present hidden  by the waiving continuum. 
Most DACs are observed with positive velocities during the first half of the orbital cycle but with negative velocities the second half (Fig.\,20). 
We notice that DACs  velocities  are scattered around the line representing the donor star velocity, and are profusely  observed at blue and red wings around \op 0.4. This is illustrated in the asymmetric distribution observed in DAC  radial velocities  (Fig.\,20). The corresponding  distribution of velocities relative to the donor is smooth and relatively symmetric. 

Interestingly, we noticed that there is no indication of the triplet structure in any of the DACs. In addition, 20\% of the DACs are double, with separation between the components of about 45 \kms. DACs have velocities a few hundreds of km s$^{-1}$ relative to the binary center of mass and are detected at low and high state.
It is notable that profiles taken 69 days apart, at $\Phi_{o}$ = 0.68 and $\Phi_{l}$ =  0.50 (HJD 2454672.52566 and  2454703.49230)  show  roughly the same set of DACs at very similar velocities.  
At high state discrete emission features  appear in the spectra and the identification of DACs is difficult. One is tempted to say that some DACs are replaced by discrete {\it emission} features at the high state.

Stationary or almost-stationary DACs  are observed in the red wing of the line, with velocities 265 and 365 \kms.   These features  could correspond to Fe\,II\,7780.4 and Fe\,II\,7783.0 with velocities 12 \kms relative to the barycenter,
but we cannot discard O\,I ``satellite'' lines (Fig.\,20).  Around \op 0.2-0.3 the DAC at 365 \kms  (eventually Fe\,II\,7783.0) attains the larger velocity (470  \kms relative to the O\,I\,7773 center) and is wider than in other epochs.  In contrast to these almost-stationary features,
we notice that some DACs seems to follow, at least at some epochs, quasi-sinusoidal patterns following the donor orbital motion.

    
The RV of the O\,I\,7773 CA presents a slight orbital oscillation with half-amplitude 32 \kms in anti-phase with the donor and center displaced by 5 \kms from  the systemic velocity. This behavior changes drastically at long cycle maximum, as seen in Fig.\,20, adopting a saw-teeth pattern, with a large velocity (+125 km s$^{-1}$) prior to $\Phi_{o}$= 0.5  and a low velocity (-125 km s$^{-1}$) just after. 
Difference profiles shown in Fig.\,21 indicate  that this behavior  is the consequence of  the disappearance of  the red (blue) emission wing during \op 0.45 (0.59). This phenomenon can be interpreted as the occultation of wind high-velocity regions by the gainer or the disc. 


 We also find DACs  in Si\,II \,6347/6371. The 50\% of the DACs detected in Si\,II\,6371 are also detected in Si\,II\,6347.  In addition,
44\% of the Si\,II DACs  have RV correspondence with those observed concurrently in O\,I\,7773. 
We find relatively strong stationary lines at 6375.8 \AA\, and 6379.2 \AA\, that we identified with diffuse interstellar bands  (Herbig 1975).

Radial velocities of broad donor-subtracted Si\,II 6347/6371 line profiles present large scatter, but they are roughly in anti-phase with the donor RVs revealing their origin in or around the gainer.


\subsection{Enhanced absorption wings}

Enhanced  absorption wings are observed in several lines, 
especially O\,I\,7773  (Fig.\,19, upper panel) and He\,I\,5875  (not shown). We measured the velocity of the wing edge with respect to the line center resulting in averages of 350 $\pm$ 25 (standard deviation) and -350 $\pm$ 20 \kms for the red and blue wing of  O\,I\,7773 and 410 $\pm$ 55 and -390 $\pm$ 50 \kms for the red and blue wing of He\,I\,5875, respectively  (Fig.\,22). 
We propose that enhanced absorption wings are caused by photon absorption in the gas stream and circumprimary disc.

  Blue depressed wings occur in the range \op 0.5-0.9, roughly coinciding with the phases when the stream faces the observer, 
and  red enhanced wings in the range \op 0.9-0.45,
just when the observer see the receding  stream (the exceptions are the blue wings around \op 0.15 and the O\,I red enhanced wings already at \op 0.85). The blue wing at \op 0.15 could be explained if part of the stream bounces back after impacting the disc producing blue-shifted additional absorption.
This behavior is similar but not strictly the same that the one followed by DACs.  In particular, the red enhanced wing appears around \op 0.9 but at this phase DACs still  are present in the blue wing. Another interesting result is the abrupt change between positive and negative velocity, with no smooth transition between them.


\begin{figure*}
\scalebox{1}[1]{\includegraphics[angle=0,width=18cm]{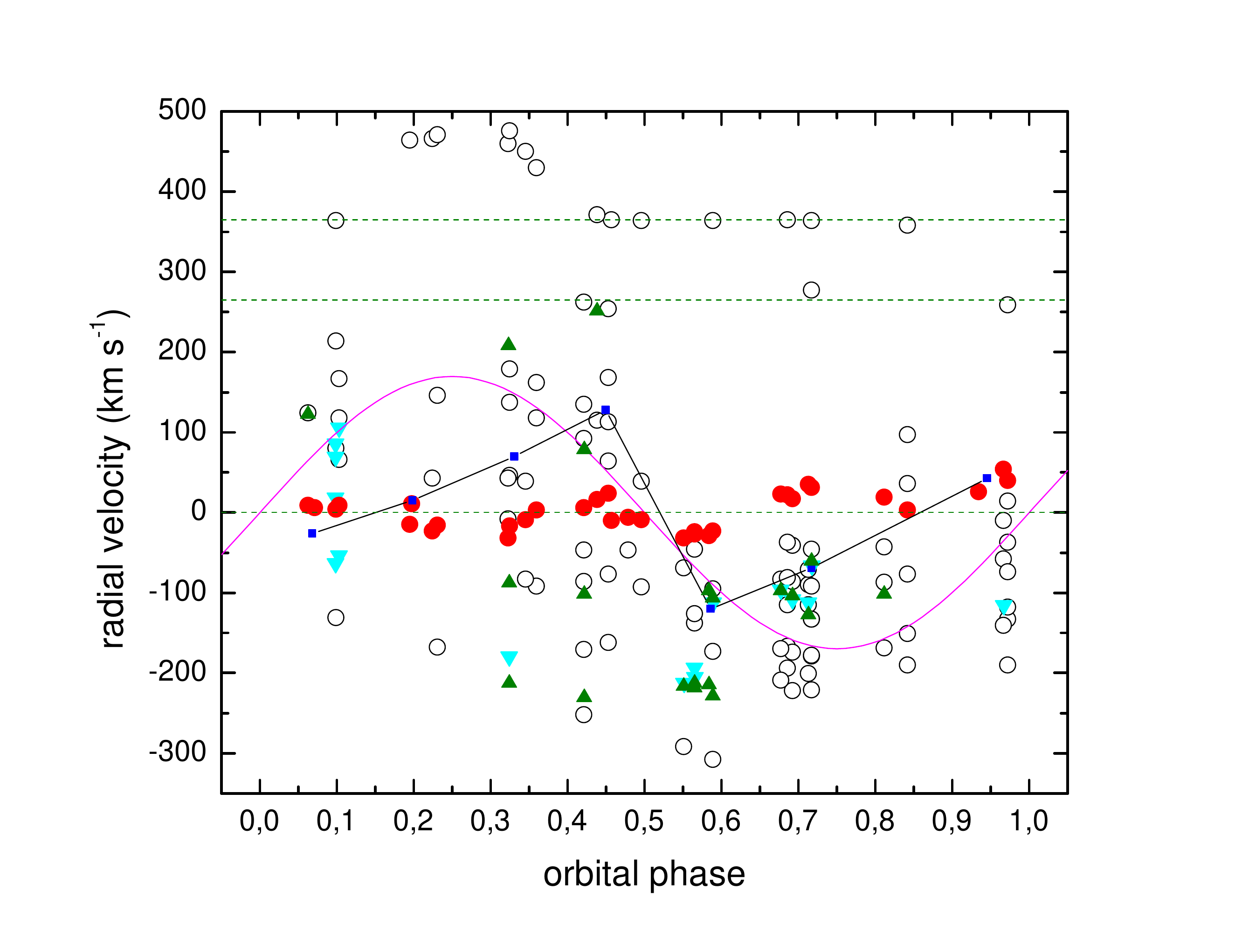}}
\caption{ Radial velocities refereed to the binary center of mass  for DACs (circles) and line center (filled circles for 0.39 $<$  $\Phi_{l}$  $<$ 0.76, filled squares for  0.05 $<$ $\Phi_{l}$ $<$ 0.07) for the O\,I\,7773 line in UVES spectra. Measurements for Si\,II\,6347 DACs (triangles) and Si\,II\,6371 DACs (inverted triangles) are also shown.  The theoretical radial velocity for the donor is shown by a sinusoid and dashed lines of constant velocity 0, 265 and 365  \kms are also displayed.}
  \label{x}
\end{figure*}



\begin{table*}
\centering
 \caption{Summary of spectroscopic variability, possible line forming regions and  line formation mechanisms. 
 ABS, CA, EM, RV and SIN stand for absorption, central absorption, emission, radial velocity and sinusoidal, respectively. Bound-bound emission is labeled b-b.} 
 \begin{tabular}{@{}lclc@{}}
 \hline
{\rm Feature}  &{\rm Variability} &  {\rm Main Region}  &Mechanism\\
 \hline
H\,I, He\,I, Si\,II \& O\,I low velocity EM &Larger at high-state &wind &b-b EM\\
H\,I and He\,I CA depth &Larger at low-state & disc &Self-ABS\\
H\,I, He\,I, Si\,II \& O\,I CA RV &Quasi-SIN, low amplitude &  disc &Self-ABS\\
Balmer $V/R$ curve &SIN $+$ perturbations on eclipses & wind   $+$ disc &Doppler $+$ Eclipses\\
He\,I V \& R peak RVs & V quasi-SIN, R stationary and shifts at high state & wind &b-b EM\\
DACs & Mostly in depressed blue wings on 2nd half-cycle & stream? &  unknown \\
Balmer extended EM wings & Red side 1st half-cycle, blue side 2nd half-cycle &wind &photon scattering \\
Enhanced ABS wings & Complex, mostly blue side 2nd half-cycle &   stream + disc  &ABS\\
\hline
\end{tabular}
\end{table*}

\section{Discussion}

Many of the phenomena described in this paper 
can be understood in terms of a global mass-loss process involving the transfer of mass between the donor and the gainer. 
In the following we discuss the spectroscopic variability in terms of 
possible line forming regions, line forming mechanisms  and mass loss/exchange channels. A comprehensive summary of all these issues is given in Table 7.

\subsection{The wind revealed in the helium emission}

It is unusual to find in a binary star stationary spectral features,   since most of the emitting regions should bring the fingerprints of the orbital motion. Therefore the discovery of stationary red emission in He\,I\,6678 is important to constrain the physical scenario. The first thing we notice is that the RV variability  of the emission peaks cannot be attributed to the
motion of the central absorption feature, as in the case of H$\alpha$. This feature is  in the middle of both peaks and seems not influence the RV of the stationary red peak.  The more simple explanation for this behavior is assuming an almost  stationary broad single emission peak divided by the CA, and that this CA has an important blue wing component  perturbing the measurement of the blue emission peak, especially at  \op 0.40, when the perturbation is large. The blue depressed wing could be due to  the gas stream facing the observer at this phase and be the low latitude manifestation of the wind producing the broad helium emission.  This wind could be produced by stream deflection at the disc/stream interaction region.  Following the same reasoning that  for Balmer lines, the helium half-peak separation (up to 450 \kms) is also consistent with a critically rotating gainer unable to accrete more material.

In principle the helium emission 
wings could be due to scattering and true absorption both in 
the line and continuum by the Schuster's mechanism (e.g. Gebbie \& Thomas 1968, Israelian \& Nikoghossian, 1996). However, it is difficult to explain
the asymmetric behavior between blue/red peak under this hypothesis.



The presence of  low velocity emission in He\,I lines confirms our previous view that the wind also emits in helium. As happens for Balmer lines, larger helium emission during the high state implies an increase in wind emissivity during this epoch. 


van Rensbergen et al. (2008) claim that gainer's spin-up and hot spots can drive mass out of a binary. They argue that 
releasing of accretion energy produces  enough energy to sustain a wind from a small stream impact region. Their model gives a physical basis for a  mass loss larger than the one driven by winds of isolated single stellar components of V\,393\,Sco. In addition, the discovery of a chromospherically active donor suggests that magnetism could play a role as a wind driver mechanism.  The symmetry of the binary configuration regarding the orbital plane indicates that this wind should be bipolar.

\begin{figure}
\scalebox{1}[1]{\includegraphics[angle=0,width=8cm]{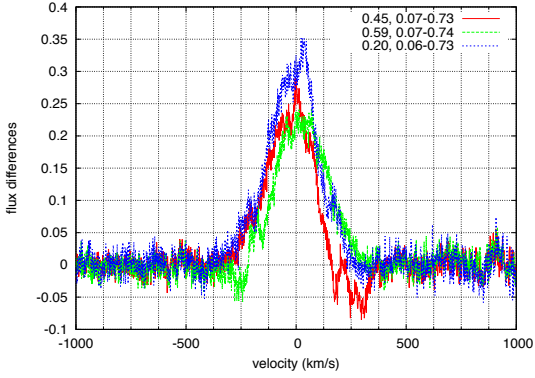}}
\caption{ Donor-subtracted OI\,7773 difference spectra during epochs of maximum RV shift at long maximum (\op 0.45 and 0.59) and zero RV shift (\op 0.20). 
Labels indicate orbital phase and the difference between the long-cycle phases. }
  \label{x}
\end{figure}

\begin{figure}
\scalebox{1}[1]{\includegraphics[angle=0,width=8cm]{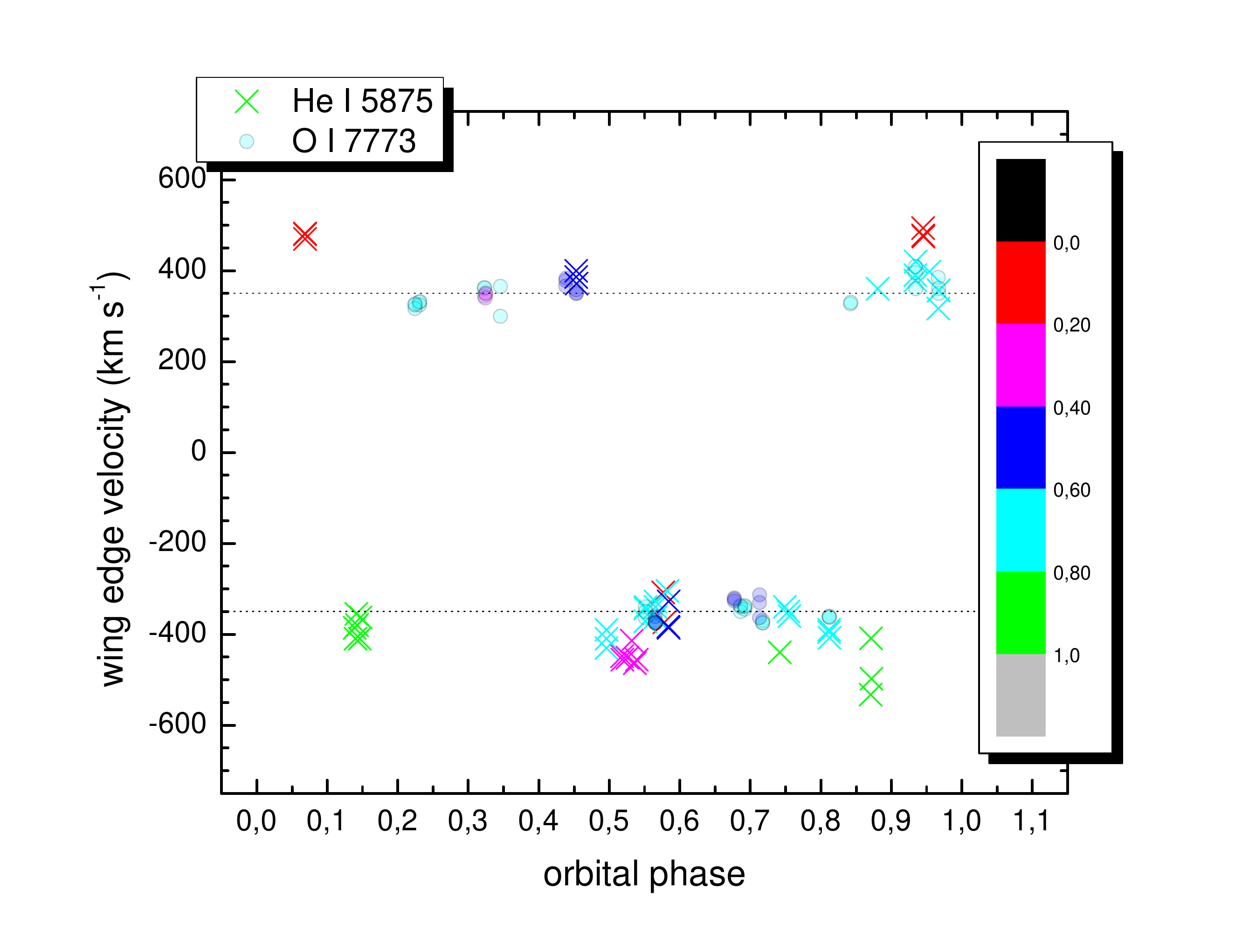}}
\caption{The velocity for the edge of the enhanced wing for O\,I\,7773 and He\,I\,5875  relative to the line center. Averages values for O\,I\,7773 are shown by dotted lines. Color bands indicate $\Phi_{l}$ ranges.}
  \label{x}
\end{figure}

\subsection{Balmer extended emission wings and the stream/disc interaction}

The presence of extended emission wings (EEWs)  in some Balmer lines beyond $\pm$ 2000 \kms cannot be attributable to emission from a Keplerian disc.  We notice that the wind projected velocities (hundreds of \kms) are too low to explain EEWs by electronic recombination,   and that the mechanism of Thompson scattering could produce extended emission wings as occurs in Be stars (Poeckert \& Marlborough 1979) and the luminous blue variable AG\,Car  (Schulte-Ladbeck et  al. 1994). 
EEWs appear in the red wing on the 1st half-cycle and in the blue wing on the 2nd half-cycle so the forming  medium shares the same orbital variability that  the medium responsible for DACs and enhanced absorption wings. However, we notice that the emission features require a medium at latitudes higher than absorbing material. 
This suggests that part of the stream 
is deflected by the disc towards high latitudes producing high density regions responsible for the Thompson scattering and EEWs. All of this activity occurs principally at the 1st and 2nd binary quadrant.

\subsection{Optically thick disc, hot spot and gas stream}

The central absorption cores of the lines studied in this paper are  probably produced in the gainer pseudo-photosphere, modeled by M12 as an optically-thick disc. The temperature inferred 
for the disc from spectroscopic line ratios $T_{hot}$ = 16.000 $K$ coincides with those inferred from the light curve analysis.
The orbital behavior of line cores is consistent with light-centers placed somewhere along the line joining the stellar centers near the gainer. Variations in the depth of the CAs
are explained by changing amounts of  filling emission as the wind increases its strength at high state. 

The gas stream is revealed at \op 0.9 in the He\,I $FWHM$ maximum  and in the asymmetries observed in O\,I and He\,I lines (Section 3.11). The hot spot,  on the other hand, shows up in the ``z-wave'' of the H$\alpha$ $V/R$ variability. 


\subsection{Line emission: the optically thin high latitude wind and donor activity} 

The increase of line emissivity during eclipses and the sustained emission increase at low velocities during high state strongly suggest that most of the low velocity  Balmer emission comes from electronic recombination at the high latitude wind. The same occurs for the emission seen in O\,I\,7773 and  the Si\,II doublet.  As discussed in Section 3.7.2 part of the H$\alpha$ emission also comes from the  disc outer edge and a region around the donor. It is remarkable that when the system approaches the high state the line emission increases in the wind (the low velocity region) {\it and} around the donor (Fig.\,16). This finding suggests a connection between the wind strength and the activity responsible for the line emission around the donor. This possible connection will be explored in a forthcoming paper dedicated to the study of weak emission features detected in the residual spectra. So far, Doppler maps show that the Mg\,II\,4481 and C\,I\,6588 residual emissions are formed in a chromospherically active star. This finding arises the possibility of a magnetically active secondary in V\,393\,Sco.

\subsection{The DAC/stream connection}



  We find that DACs follow the pattern of the enhanced absorption wings, so a connection between DACs and the gas stream is possible (see Sects. 3.10 and 3.11). 
The narrowness of DACs ($FWHM$ a few tens of \kms) indicates that they are formed in a different medium that the main O\,I\,7773 line.
Here we propose, tentatively, that DACs can be identified with tracers of donor mass loss. Traditionally, 
a continuum mass stream is considered the output of  Roche-lobe overflow. However, the presence of DACs could reveal a stream
composite of discrete and relatively narrow mass flows following magnetic field  lines connecting the donor with the disc.  
 In this view, DACs could be due to Zeeman splitting of absorption lines produced in streams projected against a bright continuum.  The
 presence of discrete emissions on high state could indicate a reordering of the streams, carrying material by high latitudes producing emission rather than absorption lines. The orbital distribution of DAC RVs could
be result of the complex structure of the magnetic field. 
 
 In order to support the above interpretation we notice that: (i)  Zeeman splitting has been reported  in the Si\,II doublet  of the related interacting binary $\beta$ Lyrae (Skulskij \& Plachinda 1993), (ii) the transient radio nature of V\,393 Scorpii has been interpreted in terms of a stellar coronal magnetic field $B \approx 10^{2}$ G (Stewart et al. 1989),  (iii) our discovery of a chromospherically active donor suggests a magnetic donor star and (iv) the discovery of a large and {\it stable} coronal 
loop observed in  Algol and RY Vul (Peterson et al. 2010, Richards et al. 2010) indicates that evidence for magnetically channeled streams has been found in DPV-related  interacting binaries.
From the theoretical point of view, the physical basis for the donor magnetism might be its rapid synchronous rotation,
 that might generate a dynamo-powered magnetic field. This idea has been explored and vindicated for Algols with late-type components (e.g. Sarna et al. 1998, Moss \& Tuominen 1997) and for rapidly rotating A-type stars (Featherstone et al. 2009). 
 

 

At present the nature of DACs is obscure,
but our findings provide strong observational constraints for any competent model, and suggest that spectropolarimetry could be a successful technique in the diagnostic of the feature's splitting.

\subsection{New insights on the long photometric cycle  }

In the past sections we have presented evidence for the existence of several circumstellar regions in V\,393\,Sco; an optically thick disc-like pseudo-photosphere,  a gas stream, a hot spot, 
a chromospherically active secondary star and a high-latitude wind revealed in almost stationary Balmer and He\,I emission.
We propose that all these regions and their associated phenomena indicate the presence of a  mass-loss phenomenon in V\,393\,Sco that is cyclically modulated with the long period. In this section we explore possible causes for this variability.

\subsubsection{On the possibility of a dynamically perturbed  disc}

Considering that dynamically perturbed discs are found in other close binary systems, we have analyzed the possibility that the long cycle is produced by a tilted circumprimary disc whose normal axis precesses around the rotational axis of the gainer. In this case we should observe deeper CA when the disc edge hides the gainer, and shallower CA when a larger projected disc area is observed. However, a tilted disc hypothesis presents severe problems: (i) it should produce large orbital changes in the equivalent width of the emission lines proportional to the projected area and also large changes in the peak separation at high state, which are not observed, (ii) it is difficult to imagine how the optically thick disc might emit at Balmer lines and (iii) the hypothesis is not consistent with the stability of the orbital LC solution (M12). In addition, the constancy of the $V/R$ light curve at high and low state strongly argues against large scale dynamical perturbations including disc precession or disc wrapping. Beside the previous evidence, the conclusion that the light source of the long-cycle is not eclipsed (M12) is not compatible with the perturbed disc hypothesis. 

\subsubsection{The  cyclic wind  and the long cycle}

We have found evidence that the long cycle in V\,393 Scorpii is produced by a wind injecting larger amounts of mass into the circumbinary medium at  high state. This conclusion is in agreement with the suggestion given by M12  that circumbinary material is responsible for the extra light during the high state. These authors give 9 $\times$ 10$^{-9}$ $M_{\odot}/yr$ for the mass transfer rate which means that  
6.2 $\times$ 10$^{-9}$ $M_{\odot}$ are injected into the interstellar medium  every long cycle in steady state regime.  This extra mass produces additional line emission in low projected velocities and also additional continuum light by free-free emission. In this view it is the changing wind that produces the long photometric cycle.
The simultaneous increase in H$\alpha$ line emissivity  near the donor and the gainer suggests a connection between the wind  and  the chromospherically active donor star. 
 The wind could be caused by stream deflection at the disc hotspot.
 The possibility of magnetic fields connecting the donor and the disc and eventually powering the wind deserves to be studied, especially knowing that the binary is a transient radio source (Stewart et al. 1989). 

It is possible that the final cause for the long cycle is self-modulation of the mass transfer rate. As the gainer is rotating at critical rotation it cannot accumulate the extra mass received from the donor and this mass is finally expelled by the wind. Spectroscopy shows that the high and low states in V\,393\,Sco correspond to changes in line and continuum emissivity, and probably density, of the wind, which could reflect changes in the mass transfer rate.  Let's assume that the mass transfer rate 
is modulated by the photon flux arriving from the gainer. This flux is attenuated by intervening material when the wind is stronger, 
producing less irradiation flux and lower $\dot{M_{2}}$. This produces a decrease in wind strength and subsequently larger irradiation flux, bigger $\dot{M_{2}}$ and so on. This valve  mechanisms could explain the long cycle, but theoretical work is needed to demonstrate that it is a valid mechanism and stable on the long term, beside of taking into account the observational evidence indicating a chromospherically active  rather than an irradiated donor. 

The wind discovered in \var  might be driven by the same (still unknown) physical mechanism that the wind found in $\beta$ Lyrae (Harmanec et al. 1996),  but there are differences between $\beta$ Lyrae and \var that have been pointed out by M12. These authors propose that $\beta$ Lyrae is in an earlier evolution stage than V\,393\,Sco,  and  characterized by a larger mass transfer rate.

\subsubsection{On equatorial and polar outflows}

Our earlier claim for equatorial mass flows in \var  (M10) can now be tested with our new  optical spectroscopic dataset. Our data suggest  that line asymmetries, especially those seen around \op 0.5 and previously interpreted as  equatorial mass flows through the L3 point, are part of a more complex orbital pattern. Although in this paper we have studied the asymmetries of the O\,I\,7773 and He\,I\,5875 lines and M10 those of the He\,I\,1083 nm line,  it is reasonable to assume that all these lines behave similarly during the orbital cycle. The new datasets show that the  blue depressed wing remains during  large part of the 2nd half cycle and cannot be interpreted in terms of an L3 outflow only  (Fig.\, 22). The new pattern  is best explained  in terms of an anisotropic wind, emerging near the system barycenter  in direction of the binary motion. On the other hand, the interpretation of the red depressed wing near \op 0.9 and 0.1 in terms of gas stream infall remains solid in this study, but the new observation of blue depressed wings  around \op 0.15 is puzzling. A possible explanation for this feature is detection of material bouncing back from the stream/disc impact region.   

M10 favored equatorial mass flows as the cause for the long cycle in part due to the absence of long cycle variability in UV superionized lines.
The resonance lines of Si\,IV, C\,IV and N\,V do not seem to be associated to the long-cycle phenomenon. It is then possible that these lines are formed in the stellar wind emerging from polar regions of the B-type gainer  (as revealed by line asymmetries discussed by M10), but probably this {\it polar-wind} is not related to the {\it disc-wind} discussed in this paper, that is clearly involved in the long-cycle variability. The IUE observations are limited to only a few spectra, this unfortunate constraint impedes to have a clear picture of superheated regions during the long-cycle.

\section{Conclusions}

In this paper we have analyzed high resolution optical spectroscopy of V\,393 Sco with time resolution suitable for  studying independently the orbital and long cycle variability. In general, the spectroscopic variability can be explained in terms of several circumstellar zones including  a circumprimary disc, a gas stream, a hot spot and a modulated high latitude wind. We have shown that the long cycle is the manifestation of this variable wind, probably produced by the interaction of the gas stream and the optically thick disc surrounding the critically rotating B-type primary.
Our main conclusions are:\\

\begin{itemize}
\item We detected the donor in our spectra; its effective temperature was determined from a comparison with theoretical synthetic spectra, obtaining $T_{2}$ = 7900 $K$. The donor RV was measured and the orbital elements determined; they are given in Table\,5. We find a significant small orbital eccentricity for the system, but its meaning is not clear in an interacting binary with mass flows.
\item From the strength of diffuse interstellar bands we estimate $E(B-V)$ = 0.15 $\pm$ 0.05, consistent with the M10 result.
\item We developed a method of spectral disentangling subtracting  from every spectrum at every observing epoch the donor synthetic spectrum weighted by its fractional contribution given by our light curve model.
This method turned to be adequate to remove the main spectral features of the donor and for studying the remaining components of the system.
\item At low stage the He\,4471 and Mg\,4481 equivalent widths are consistent with line formation in an optically thick region with temperature 16.000 $K$. We argue that at low state these lines are less contaminated by emission and probe the pseudo-photosphere of the B-type gainer.
\item Residual spectra reveal several metallic emission lines from the donor, like Mg\,II\,4481 and C\,I\,6588. Doppler maps of these lines reveal a homogeneously-distributed emissivity around the secondary, not compatible with irradiation-induced emission but with a chromospherically active star. The highly variable nature of this emission along with the reported radio-transient  nature of the binary rise the question on the probable magnetic nature of the donor.
\item Novel spectral features were detected: discrete absorption components (DACs), especially visible at blue-depressed O\,I\,7773 absorption wings during the second half-cycle, Balmer double emission with $V/R$ curves showing ``z-type`` and ``s-type'' excursions around secondary and main eclipse, respectively, and H$\beta$ emission wings extending up to $\pm$ 2000 km s$^{-1}$. 
\item  DACs are a new and still enigmatic phenomenon;  deeper studies  are needed to enlightening the nature of these features. We suggest that DACs could be Zeeman separated absorption lines formed in magnetized gas streams. This conjecture can be tested with spectropolarimetry.
\item  Line asymmetries can be interpreted  in terms of   photon absorption in the canonical gas stream, a large-scale wind and material bouncing back from the stream/disc impact region.
\item We notice that the relatively large widths of the Balmer and helium absorption cores are consistent with rotational broadening in absorbing  material around a critically rotating gainer.
\item Balmer and He\,I emission are not eclipsed and in average the 
H$\alpha$ violet peak is larger than the red one. At phases of maximum disc visibility and on high state the emission profiles are single-peaked. These findings  suggest that Balmer and He\,I emission are produced  in a high-latitude bipolar wind. 
\item The extended Balmer emission wings can be the result of Thompson scattering in dense regions  of this wind. These regions are in the binary hemisphere containing  the stream-disc interaction region  and could be formed by stream deflection at the hot spot.
\item As cause for the long cycle we exclude a dynamical  instability like disc precession or disc wrapping. This conclusion is based on the stability of the $V/R$ curve, the stability of the photometric orbital light curve, the variability of spectral features like peak separation and most importantly the fact that the long cycle is produced by a non-eclipsed source.
\item We interpret the long-cycle in terms of a  bipolar wind whose emissivity is modulated with the long period. This wind produces rather stationary Balmer, O\,I, Si\,II and He\,I broad emissions.  On  high state the wind strength is larger than on  low state. We argue that an increase in wind strength is the cause for the larger line emission and  brighter and redder continuum  on the high state.
\item It is very likely that the wind is the result of the interaction between the gas stream and the optically thick and massive circumprimary disc. A donor magnetic field could play an important role in the energetic of the  phenomenon and acting as a driving mechanism for the wind. 

 \end{itemize}





\section{Acknowledgments}
 We acknowledge an anonymous referee for valuable comments that helped to improve the first version of this manuscript.
REM acknowledges support by Fondecyt grant 1070705, 1110347, the Chilean 
Center for Astrophysics FONDAP 15010003 and  from the BASAL
Centro de Astrof\'isica y Tecnologias Afines (CATA) PFB--06/2007. 
G. D. gratefully acknowledge the financial support of the Ministry of Education and Science of the Republic of Serbia through the project 176004, ÓStellar physicsÓ.



\bsp 
\label{lastpage}
\end{document}